\documentclass[twocolumn]{aastex7}

\usepackage{enumitem}
\usepackage{amsmath}
\usepackage{latexsym}
\usepackage{graphicx}
\usepackage{epsf}
\usepackage{CJK}
\usepackage{wasysym}
\usepackage{natbib}
\usepackage{physics}
\usepackage[T1]{fontenc}

\newcommand{\oiii}{[O\,{\sc iii}]}

\newcommand{\hei}{He\,{\sc i}}
\newcommand{\ha}{H$\alpha$}

\newcommand{\lya}{Ly$\alpha$}

\newcommand{\utoirac}{$u$-to-IRAC}

\received{MMDDYY}
\revised{MMDDYY}
\accepted{MMDDYY}

\submitjournal{ApJL}

\shorttitle{Spectroscopic Sample of Cosmic Noon LRDs}
\shortauthors{Y. Ma et al}

\graphicspath{{./}{figures/}}

\begin{document}
\begin{CJK*}{UTF8}{gbsn}
\title{Little Red Dots at Cosmic Noon: \\Spectroscopy Reveals Declining Abundance but a Non-Evolving Luminosity Function Shape}

\correspondingauthor{Yilun Ma (yilun@princeton.edu)}
\author[0000-0002-0463-9528]{Yilun Ma (马逸伦)}
\affiliation{Department of Astrophysical Sciences, Princeton University, Princeton, NJ 08544, USA}
\email{yilun@princeton.edu}

\author[0000-0002-7847-1107]{Daniel J. Lyon}
\affiliation{Centre for Astrophysics and Supercomputing, Swinburne University of Technology, Melbourne, VIC 3122, Australia}
\email{}

\author[0000-0002-5612-3427]{Jenny E. Greene}
\affiliation{Department of Astrophysical Sciences, Princeton University, Princeton, NJ 08544, USA}
\email{jgreene@astro.princeton.edu}

\author[0000-0001-5586-6950]{Alberto Torralba}
\affiliation{Institute of Science and Technology Austria (ISTA), Am Campus 1, 3400 Klosterneuburg, Austria}
\email{alberto.torralba@ista.ac.at}

\author[0000-0003-2871-127X]{Jorryt Matthee} 
\affiliation{Institute of Science and Technology Austria (ISTA), Am Campus 1, 3400 Klosterneuburg, Austria}
\email{jorryt.matthee@ist.ac.at} 

\author[0000-0002-2057-5376]{Ivo Labbe}
\affiliation{Centre for Astrophysics and Supercomputing, Swinburne University of Technology, Melbourne, VIC 3122, Australia}
\email{ilabbe@swin.edu.au}

\author[0000-0003-4700-663X]{Andy D. Goulding} 
\affiliation{Department of Astrophysical Sciences, Princeton University, Princeton, NJ 08544, USA}
\email{goulding@astro.princeton.edu}

\author[0000-0003-4075-7393]{David J. Setton} 
\affiliation{Department of Astrophysical Sciences, Princeton University, Princeton, NJ 08544, USA}
\email{davidsetton@jhu.edu}


\author[0000-0002-8053-8040]{Marianna Annunziatella}
\affiliation{INAF-IASF Milano, via A. Corti 12, I-20133 Milano, Italy}
\email{mannunziatella@cab.inta-csic.es}

\author[0009-0008-9877-5512]{Aayush Desai}
\affiliation{Institute of Science and Technology Austria (ISTA), Am Campus 1, 3400 Klosterneuburg, Austria}
\email{aayush.desai@ist.ac.at}

\author[0000-0003-1344-9475]{Eiichi Egami} 
\affiliation{Steward Observatory, University of Arizona, 933 N Cherry Avenue, Tucson, AZ 85721, USA}
\email{egami@arizona.edu}

\author[0000-0003-3310-0131]{Xiaohui Fan}
\affiliation{Steward Observatory, University of Arizona, 933 N Cherry Avenue, Tucson, AZ 85721, USA}
\email{xiaohuidominicfan@gmail.com}

\author[orcid=0000-0002-3254-9044]{Karl Glazebrook} 
\affiliation{Centre for Astrophysics and Supercomputing, Swinburne University of Technology, Melbourne, VIC 3122, Australia}
\email{kglazebrook@swin.edu.au}  

\author[0000-0002-5588-9156]{Vasily Kokorev}
\affiliation{Department of Astronomy, The University of Texas at Austin, Austin, TX 78712, USA}
\email{vkokorev@utexas.edu}

\author[0000-0001-9592-4190]{Jiaxuan Li (李嘉轩)}
\email{jiaxuanl@stanford.edu}
\affiliation{Kavli Institute for Particle Astrophysics and Cosmology, Stanford University, Stanford, CA 94305, USA}

\author[0000-0001-6052-4234]{Xiaojing Lin}
\affiliation{Department of Astronomy, Tsinghua University, Beijing 100084, People's Republic of China}
\email{xiaojinglin.astro@gmail.com}

\author[0000-0001-9002-3502]{Danilo Marchesini}
\affiliation{Department of Physics and Astronomy, Tufts University, Medford, MA 02155, USA}
\email{danilo.marchesini@tufts.edu}

\author[orcid=0000-0003-2804-0648,sname='Themiya Nanayakkara']{Themiya~Nanayakkara}
\affiliation{Sydney Institute for Astronomy, School of Physics A28, The University of Sydney, NSW 2006, Australia.}
\email[]{themiya.nanayakkara@sydney.edu.au}

\author[0000-0002-6265-2675]{Luke Robbins}
\affiliation{Department of Physics and Astronomy, Tufts University, Medford, MA 02155, USA}
\email{andrew.robbins@tufts.edu}

\author[0000-0002-1917-1200]{Anna Sajina}
\affiliation{Department of Physics and Astronomy, Tufts University, Medford, MA 02155, USA}
\email{annie.sajina@gmail.com}

\author[0000-0002-7712-7857]{Marcin Sawicki}
\affiliation{Institute for Computational Astrophysics and Department of Astronomy and Physics, Saint Mary's University, 923 Robie Street, Halifax, Nova Scotia, B3H 3C3, Canada}
\email{marcin.sawicki@smu.ca}
\affiliation{NRC-Herzberg, 5071 West Saanich Road, Victoria, British Columbia V9E 2E7, Canada}

\begin{abstract}

Little red dots (LRDs) are an extremely abundant population of red, compact, broad-line objects primarily found in the early universe. At $5<z<8$, they are at least 10 times more abundant than UV-selected quasars. Yet, how this population evolves towards lower redshift is unclear. In this work, we present our search for LRDs at cosmic noon ($2\lesssim z\lesssim 4$) in nearly 20\,deg$^2$ of the Deep/UltraDeep fields of the Hyper Suprime-Cam Subaru Strategic Program (HSC-SSP). Combining multiwavelength wide-field imaging surveys from 0.3 to 4.5\,$\mu$m, we identify 266 photometric LRD candidates. With spectroscopic follow-up from Keck/NIRES, VLT/X-shooter, and Magellan/FIRE on 24 candidates, we identify a sample of four LRDs with broad \ha\ emission. With the purity of the photometric sample calibrated by spectroscopy, we estimate the number density of LRDs to be approximately $6.8\times10^{-7}\,\mathrm{cMpc^{-3}}$ at $2.10<z<2.55$---this is nearly a factor of 35 decrease from their flat number density evolution at $4\lesssim z\lesssim8$ estimated from JWST spectroscopic surveys. The knee of the LRD optical luminosity function also remains consistently at $M_{5500}\approx-22$\,mag across $2<z<5$, below that of quasars. We further speculate that only the abundance of LRDs, but not the shape of their optical luminosity function, evolves with redshift, and the Roman Space Telescope will be able to test this decisively with a substantially larger sample of low-redshift LRDs.\\




\end{abstract}

\keywords{\uat{Active galactic nuclei}{16}, \uat{Black holes}{162}, \uat{Galaxy formation}{595}, \uat{High-redshift galaxies}{734}}

\section{Introduction}\label{sec:intro}

Little red dots (LRDs; \citealt{Matthee2024}) are arguably one of the most surprising discoveries of the James Webb Space Telescope (JWST) in the early universe. They exhibit compact morphology ($r_\mathrm{e}\lesssim200\,\mathrm{pc}$) and a characteristic V-shaped spectral energy distribution (SEDs) consisting of a faint, blue UV component and an extremely red optical continuum \citep{Kokorev2024, Akins2025, Kocevski2025, Labbe2025}. They often show Balmer emission lines broadened to $\sim$2000\,$\mathrm{km\,s^{-1}}$ \citep{Greene2024} suggesting active galactic nucleus (AGN) activity. However, they do not resemble typical type-1 AGN in many ways. No strong X-ray emission has been detected in LRDs \citep{Ananna2024, Yue2024}. Despite their red optical color, they also lack thermal emission from the typical AGN dusty torus in the near-infrared or from obscured star formation activity in the far-infrared \citep{Setton2025alma, Xiao2025noema}. Variability has only been tentatively detected in strongly lensed and multiply imaged LRDs on a 100\,year-timescale \citep{Zhang2025lensvariability}. Furthermore, some LRDs also show Balmer breaks that are much stronger than what stellar populations alone could produce \citep{Furtak2024qso1, Labbe2024monster, deGraaff2025cliff, Ma2025qso1, Wang2025brd, Naidu2026bhstar}. This motivates a class of models where the central engine of LRDs is enshrouded in an envelope of dense gas, whose photosphere emits like a blackbody with $T_\mathrm{eff}\sim3000$--7000\,K resulting in the intrinsic redness and the ubiquitous, strong Balmer break of LRDs \citep{InayoshiMaiolino2025, Liu2025break}. Models of supermassive stars have also been proposed to explain these spectral features given their similarity to stellar objects \citep{Chisholm2026gc, Lu2026sms, NandalLoeb2026}. Yet, so far, no model has been able to consistently explain the unique combination of features observed in LRDs. 

In addition to their peculiar SED shape, the demographics of LRDs is also puzzling. At $z\gtrsim4$, they are at least 10 times more abundant than UV-selected quasars if extrapolating the quasar luminosity function to LRD luminosity---a result that has been consistently confirmed \citep{Greene2024, Kokorev2024, Akins2025, Kocevski2025}. At $z\gtrsim4$, they also exhibit a sharp luminosity cutoff at $M_\mathrm{5100\,\AA}\sim -23$ (or $L_\mathrm{5100\,\AA}\sim10^{44.5}\,\mathrm{erg\,s^{-1}}$), which may suggest an upper bound in black hole mass at $M_\mathrm{BH}\sim10^7\,M_\odot$ \citep{Ma2026cutoff}. However, at cosmic noon, measurements of LRD abundances are more uncertain. On one hand, early LRD searches with JWST photometry found a disappearance of LRDs from $z\sim5$ to $z\sim2$ \citep{Kocevski2025}. Yet, these JWST searches lack robust optical wavelength coverage (although F814W imaging from the Hubble Space Telescope was used). That coverage is required to select for the blue UV continuum of LRDs, an essential component of their V-shaped SEDs; thus, selection effects could account for this disappearance. On the other hand, spectroscopic searches of LRDs at cosmic noon with JWST do not provide samples large enough (e.g., $\sim5$ objects in \citealt{Kapoor2026z2}) to robustly pin down LRD abundances because of the small field of view of JWST; they also often lack \ha\ spectroscopy to confidently identify the LRD nature of each object \citep{Kapoor2026z2, Loiacono2026}. The abundance measurements of LRDs at cosmic noon estimated using these JWST spectroscopic samples often show differences of up to 1\,dex, partially due to small-number statistics. 

The community has thus sought to overcome these challenges by identifying LRDs in wide-field imaging surveys \citep{Bisigello2026, Ma2026z2}, which provide uniform optical and near-infrared coverage corresponding to rest-frame UV and optical wavelengths at $z\sim2$--3, over areas substantially larger than those surveyed by JWST. However, photometrically selected samples suffer contamination from sources such as red galaxies and dust-obscured AGN, which can mimic the V-shaped SEDs of LRDs \citep{Ma2026z2}. Differences in sample purity can lead to a scatter of $\gtrsim1$\,dex among LRD abundance estimations from different wide-field surveys that invoke different selection criteria. Spectroscopic follow-up observations of candidate sources selected from wide-field imaging surveys are necessary to determine the purity of photometric samples. 

In \citet{Ma2026z2}, our team conducted a pilot search for LRDs at $2\lesssim z\lesssim 4$ over 3\,deg$^2$ of the UltraDeep fields of the Hyper Suprime-Cam Subaru Strategic Program (HSC-SSP; \citealt{Aihara2018hscssp}), finding a steep decline (by nearly a factor of 20) in their abundance from $z\sim5$ to $z\sim2$. As a proof of concept, we also presented our first successful spectroscopic confirmation of an LRD from this photometric sample, DEEP23-z2LRD1 at $z=2.26$, whose $K$-band spectrum reveals broad \ha\ emission. However, with coverage of 3\,deg$^2$, the pilot sample only has robust constraining power on the normalization but not the shape of LRD's luminosity function. Moreover, a single spectroscopic confirmation is insufficient to establish the purity of the photometric selection. A larger spectroscopic campaign targeting LRD candidates selected over an even wider area is thus essential to robustly constrain the luminosity function of LRDs at cosmic noon. 

In this paper, we present our search of LRDs at cosmic noon within nearly 20\,deg$^2$ of the four Deep/UltraDeep fields of the HSC-SSP and the spectroscopic follow-up campaign targeting a subset of these identified LRD candidates. The paper is structured as follows. Section~\ref{sec:data} gives an overview of the parent multiwavelength catalog from which the LRDs are selected from. Section~\ref{sec:sample_selection} describes the selection methods. Section~\ref{sec:spec_followup} describes spectroscopic follow-up observations of the LRD candidates. We present our LRD number-density estimation in Section~\ref{sec:analysis} and discuss its implications in Section~\ref{sec:discussion}. Lastly, we summarize our results in Section~\ref{sec:summary}. Throughout the paper, we assume a cosmology with $H_0=70\,\mathrm{km\,s^{-1}\,Mpc^{-1}}$, $\Omega_\mathrm{m,0}=0.3$, and $\Omega_{\Lambda,0}=0.7$. All magnitudes are presented in AB magnitude \citep{Oke&Gunn1983}. 

\section{Multiwavelength Imaging Surveys}\label{sec:data}
Our search covers 19.9\,deg$^2$ across the four Deep and UltraDeep (D/UD) fields of HSC-SSP. These fields contain extensive multiwavelength imaging spanning the observed near-UV to infrared, allowing us to identify the characteristic V-shaped SEDs of LRDs. The deep uniform optical imaging from HSC reaches a 5$\sigma$ point-source detection limit of $i=26.9$ \citep{Aihara2022hscdr3}. The HSC imaging is complemented by $u/u^*$-band observations from the CFHT Large Area $U$-band Survey (CLAUDS; \citealt{Sawicki2019clauds}) across the D/UD fields. Near-infrared coverage in the $JHK/K_\mathrm{s}$ bands is from several ground-based surveys, including the Ultra Deep Survey with the VISTA Telescope (UltraVISTA; \citealt{McCracken2012uvista}), the VISTA Deep Extragalactic Observations Survey (VIDEO; \citealt{Jarvis2013video}), the Deep eXtragalactic Survey and Ultra-Deep Survey of the UKIRT Infrared Deep Sky Survey (UKIDSS/DXS and UKIDSS/UDS; \citealt{Lawrence2007ukidss}), the Deep UKIRT Near-infrared Steward Survey (DUNES$^2$; \citealt{Aihara2022hscdr3}, Egami et al., in prep.), and the DeepCos survey (PI: Y.-T. Lin; \citealt{Aihara2022hscdr3}). At longer wavelengths, we include the 3.6 and 4.5\,$\mu$m Spitzer/IRAC mosaics assembled by \citet{Lacy2021deepdrill} and the Spitzer Coverage of the HSC-Deep with IRAC for Z-studies (SHIRAZ; \citealt{Annunziatella2023shiraz}). A comprehensive description of the multiwavelength imaging available in the HSC D/UD fields is presented by \citet{Aihara2022hscdr3}, while \citet{Lacy2021deepdrill} and \citet{Annunziatella2023shiraz} provide further details on the archival observations incorporated into the IRAC mosaics used in this work.

We combine the multiwavelength imaging described above into the ``\utoirac'' photometric catalog. From the $u/u^*$ band to the $K/K_\mathrm{s}$ bands\footnote{For simplicity, we hereafter refer to $u/u^*$ and $K/K_\mathrm{s}$ as $u$ and $K$, respectively, unless otherwise specified.}, source detection and photometry are performed using a modified HSC pipeline that accommodates imaging from other facilities \citep{Bosch2018hscpipe, Bosch2019lsstpipe, Desprez2023clauds+hsc}. Sources are detected by identifying peaks simultaneously across all bands, with the detection band assigned as the first band exhibiting a $>7\sigma$ peak in the order of $irzygJHK$. Forced photometry is subsequently performed at the detected positions in all ground-based bands, following the procedures outlined in \citet{Aihara2022hscdr3}. For the Spitzer/IRAC imaging, we employ an established source deblending technique \citep[e.g.,][]{Labbe2006, Wuyts2007, Marchesini2009smf, Williams2009, Whitaker2011}. Sources are initially identified in higher-resolution $K$, $J$, or $z$-band images, preferentially adopting the reddest available band. These images are then convolved to the IRAC resolution using spatially varying kernels derived from bright, isolated, unsaturated sources detected in both the high-resolution images and the IRAC mosaics. The kernels are constructed by fitting their Fourier transforms with Gaussian-weighted Hermite functions. Each kernel is visually inspected, and those exhibiting poor fits or significant deviations are rejected. The coefficients of the remaining kernels are then interpolated into a smoothed two-dimensional map that captures spatial variations in the resolution matching. Using these kernels, all sources within each IRAC frame are modeled simultaneously, with their fluxes treated as free parameters. The modeled contributions from neighboring sources are subsequently subtracted to produce the deblended IRAC images, from which fluxes are measured using fixed circular apertures of $D=3''$.
We then apply aperture corrections to estimate total fluxes, using median curves of growth constructed from point sources in each field's IRAC mosaic. A comparison with the COSMOS2020 catalog \citep{Weaver2022} yields a median magnitude offset of 0.01\,mag and a scatter of approximately 0.1\,mag for sources with $19<\mathrm{AB}<23$ in both IRAC channels (L. Robbins, priv. comm.).

Lastly, we also remove objects that only have peaks detected in less than three bands and objects that are blended with bright stars using the star masks provided in the catalog. These sources are not involved in estimating the total survey area.

\section{Sample Selection}\label{sec:sample_selection}

In this section, we describe our photometric selection and its completeness estimation for our LRD search at cosmic noon using the \utoirac\ catalog. We also present how photometric redshifts are estimated from the multi-band photometry. 

\begin{figure*}
    \centering
    \includegraphics[width=\linewidth]{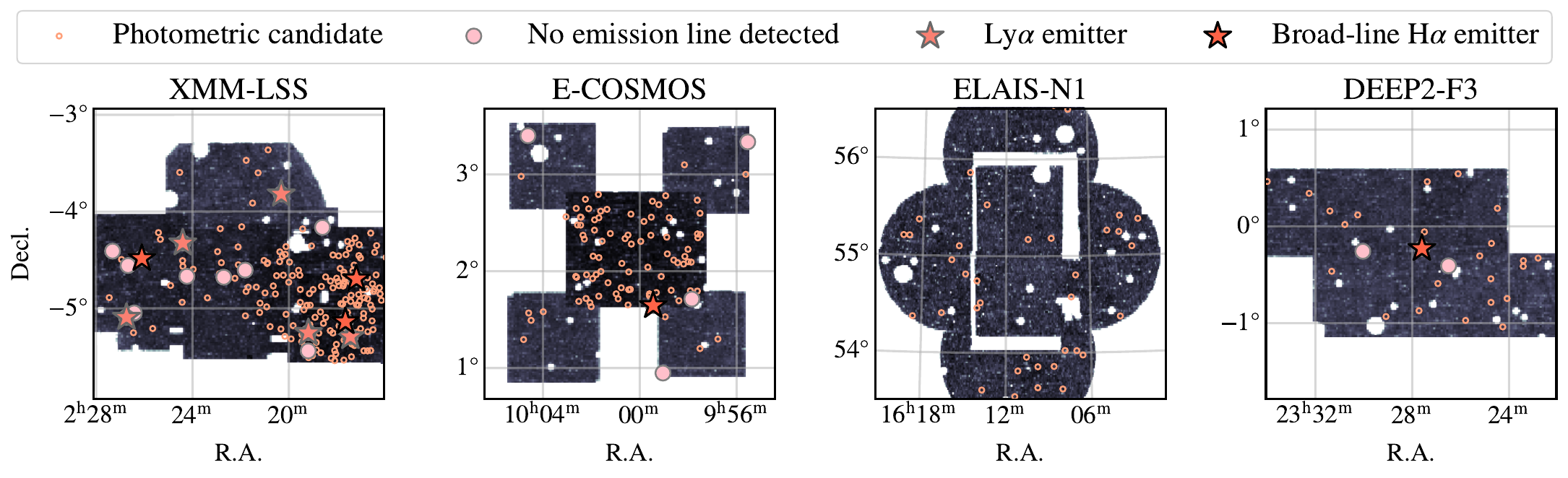}
    \caption{The photometric LRD candidates and spectroscopically targeted sources are shown in the RA--Dec plane. Dark red stars represent sources with a confirmed broad H$\alpha$ emission line; light red stars represent sources where only a spectroscopic redshift is obtained from Ly$\alpha$ emission, but no broad H$\alpha$ available and/or detected. Pink circles represent sources that were targeted spectroscopically but no spectroscopic redshift obtained. Empty circles represents photometric candidates. The backgrounds are 2D-histograms showing source densities and survey footprints of the parent \utoirac\ catalog from which the LRD candidates were selected. }
    \label{fig:RADEC}
\end{figure*}

\subsection{Photometric Selection}\label{sec:phot_selection}

\begin{figure*}
    \centering
    \includegraphics[width=\linewidth]{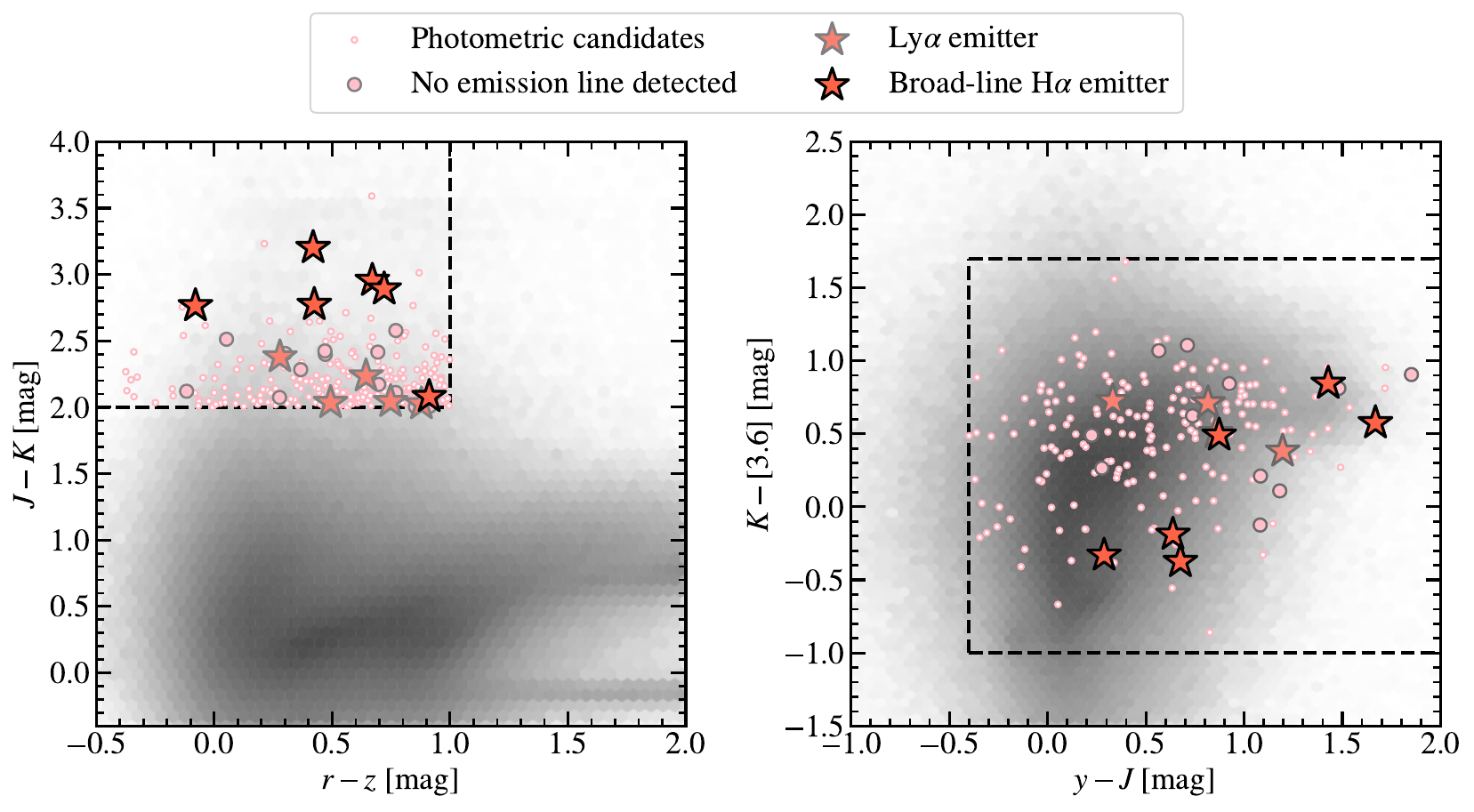}
    \caption{\textbf{Left:} The LRD candidates and spectroscopically confirmed LRDs are shown in the rest-frame UV-optical color space. Same as in Figure~\ref{fig:RADEC}, dark red stars represent sources with a confirmed broad H$\alpha$ emission line; light red stars represent sources where only a spectroscopic redshift is obtained from Ly$\alpha$ emission, but no broad H$\alpha$ is available and/or detected. Pink circles represent sources that were targeted spectroscopically but no spectroscopic redshift was obtained. Empty circles represents photometric candidates. The gray background are sources in the parent \utoirac\ catalog from which the LRDs are selected. The dashed lines mark the selection criteria. \textbf{Right:} Same as the left panel but objects are shown in the rest-frame optical-NIR color space. }
    \label{fig:colorcolor}
\end{figure*}

From the \utoirac\ catalog, we select photometric LRD candidates following the same criteria used by \citet{Ma2026z2} to identify LRDs at $1.7<z<3.7$. We outline the criteria here:
\begin{itemize}
    \item signal-to-noise ratio (SNR) is greater than 3 in all bands from $r$ to $K$;
    \item the object is a point source with \texttt{extendedness=0} (defined by $m_\mathrm{PSF}-m_\mathrm{CMODEL}<\Delta m_\mathrm{threshold}$; see \citealt{Bosch2018hscpipe}) in the $i$, $J$, $H$, and $K$ bands\footnote{The $i$-band seeing is 0.66$''$ in the HSC D/UD fields \citep{Aihara2022hscdr3}, corresponding to $\sim5.4\,\mathrm{kpc}$ at $z=2.3$.};
    \item $J-K>2$ for the steep rise in the rest-frame optical;
    \item $r-z<1$ for blue color in the rest-frame UV. 
\end{itemize}
These primary selection criteria are supplemented by two additional color cuts:
\begin{itemize}
    \item $y-J > -0.4$,
    \item $-1 < K-[3.6] < 1.7$, when \textit{Spitzer}/IRAC photometry is available.
\end{itemize}
These additional cuts are designed to reject sources with unreliable photometry and/or SED shapes inconsistent with those characteristic of LRDs; we refer the reader to \citet{Ma2026z2} for a detailed discussion. 

Using these selection criteria, we identify 341 photometric LRD candidates from the 20-deg$^2$ HSC-D/UD fields. We further conduct visual inspection on the images of these candidates to remove impostors such as nearby neighbors, streaks, and other reduction artifacts. Finally, we obtain a sample of 266 photometric LRD candidates. A breakdown of number of remaining objects after applying each criterion is provided in Appendix~\ref{app:sel_breakdown}. We show these candidates in the RA--Dec space in Figure~\ref{fig:RADEC}. We also show their positions in the color-color space in Figure~\ref{fig:colorcolor}. 

\subsection{Photometric Redshift}\label{sec:photoz}
We estimate the photometric redshift of the LRD candidates using the Python version of Easy and Accurate $z_\mathrm{phot}$ from Yale (\texttt{EAZY}; \citealt{Brammer2008eazy}) with the default \texttt{tweak\_fsps\_QSF\_12\_v3} template suites. Similar to the treatment in \cite{Ma2026z2}, this default suite is also supplemented with the UV-to-near-infrare SED of UNCOVER-45924 \citep{Labbe2024monster} and RUBIES-BLAGN-1 \citep{Wang2025brd}---two bright LRDs whose SEDs have been studied extensively in detail---to account for the H$\alpha$ emission and the red optical continuum of LRDs not easily characterized with the default templates alone. Since the equivalent width of H$\alpha$ emission from LRDs is so high that it often produces a $K$-band excess at $2.1\lesssim z\lesssim 2.5$ (see \citealt{Ma2026z2}), including LRD templates in \texttt{EAZY} allows the code to favor an LRD-dominated solution over a heavily reddened galaxy continuum at a higher redshift when such an excess is present in the photometric SED, thus improving the accuracy of the photometric redshift estimate. We show the effect of the additional LRD templates in the photo-$z$ estimation in Figure~\ref{fig:photo_z_comparison} of Appendix~\ref{app:photoz}. 

\subsection{Selection Completeness}\label{sec:completeness}

\begin{figure}[ht!]
    \centering
    \includegraphics[width=\linewidth]{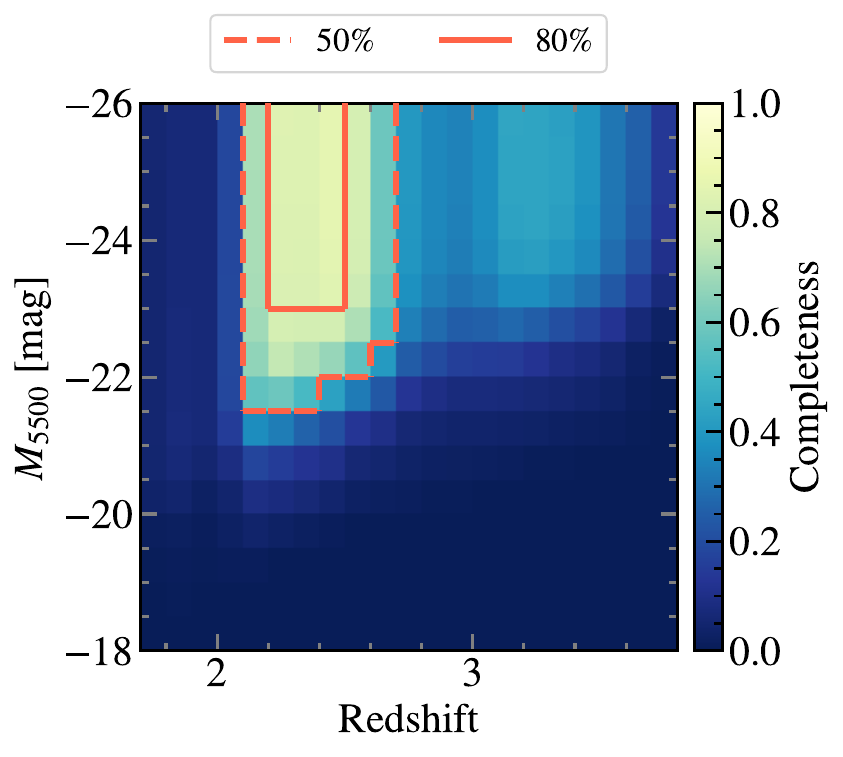}
    \caption{The selection completeness of the LRD selection as a function of optical luminosity and redshift averaged over the four HSC-D/UD fields. The lines are contours for a completeness of 50\% ($2.1<z<2.6$) and 80\% ($2.2<z<2.5$).}
    \label{fig:completeness}
\end{figure}

\begin{figure*}[ht]
    \centering
    \includegraphics[width=\linewidth]{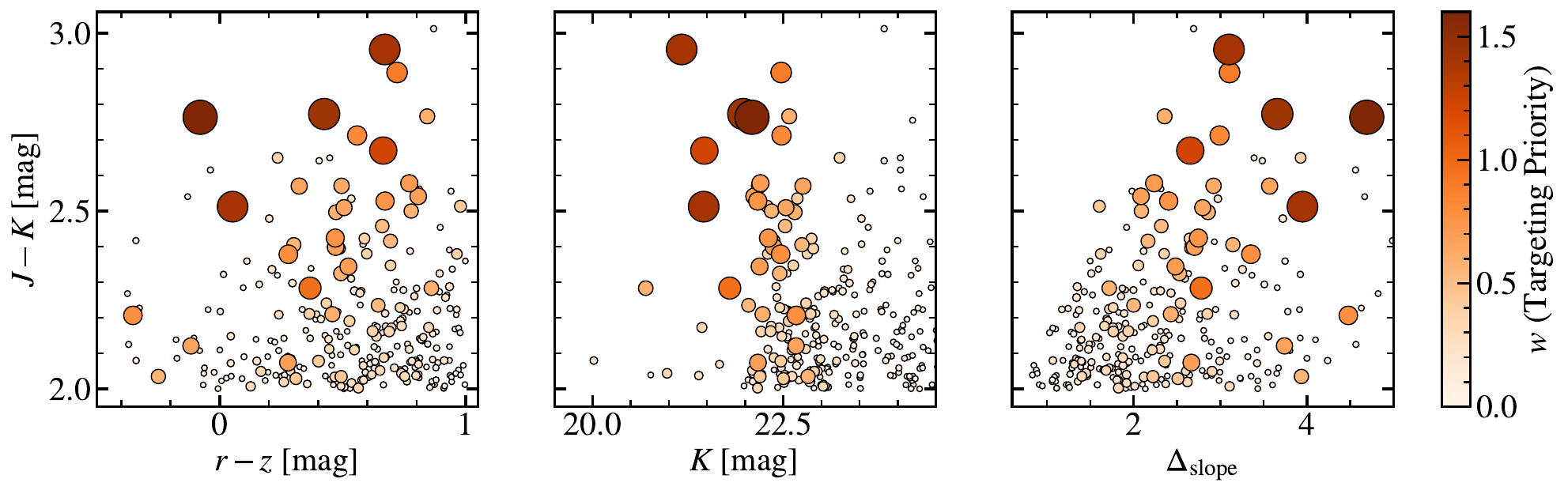}
    \caption{All sources in the parent \utoirac\ catalog are color-coded and size-coded by the targeting priority weight $w$ defined in Section~\ref{sec:target_selection}. Darker color and larger marker size represent higher targeting priority weight. The definition of $\Delta_\mathrm{slope}$, the UV-optical slope contrast is provided in Equation~\ref{eq:dslope}. }
    \label{fig:weight_visual}
\end{figure*}

Measuring the luminosity function of LRDs requires an assessment of both the completeness and the purity of our selection. We estimate the former using synthetic photometry and empirically constrain the latter through follow-up spectroscopy (see Sections~\ref{sec:spec_followup} and \ref{sec:olf}). We test the completeness of our photometric selection using 65 of the spectroscopically confirmed LRDs with high-SNR NIRSpec/PRISM spectra \citep{deGraaff2026sample}. We create a suite of synthetic \utoirac\ photometric SEDs from the PRISM spectra of these JWST-selected LRDs at a grid of optical luminosities ($M_{5500}\in[-26, -18]$ with $\Delta M=0.5$ increments) and redshifts ($z\in[1.7, 3.8]$ with $\Delta z=0.1$ increments). 

To properly generate mock observations, these synthetic SEDs need to be perturbed with photometric uncertainties. Firstly, we divide the full survey footprints into subregions defined by the overlap between the different near-infrared surveys and the varying depths of the HSC imaging. For each band within each subregion, we then identify all objects in the parent \utoirac\ catalog whose magnitudes are within $\pm0.25\,\mathrm{mag}$ of the synthetic photometry, and randomly draw a photometric uncertainty from the uncertainties of these objects. We then use these uncertainties to perturb the synthetic photometry. The perturbed synthetic SEDs are finally fed into the selection pipeline described in Section~\ref{sec:phot_selection} to estimate the completeness of our photometric selection. We show the selection completeness averaged over all four HSC-D/UD fields in Figure~\ref{fig:completeness} and in each field in Figure~\ref{fig:completeness_multi} in Appendix~\ref{app:completeness}. Our selection is complete at the $\gtrsim50\%$-level for sources at $2.1<z<2.6$ and reaches $\gtrsim80\%$-completeness for sources at $2.2<z<2.5$. This is due to the $K/K_\mathrm{s}$ excess originating from the high-equivalent-width (EW) \ha\ emission of LRDs. 

\section{Spectroscopic Follow-up}\label{sec:spec_followup}
We obtained (optical-to-)NIR spectra of 24 LRD candidates in the XMM-LSS, E-COSMOS, and DEEP2-F3 fields using the Near-Infrared Echellette Spectrometer (NIRES; \citealt{Wilson2004nires}) on the Keck II telescope, the X-shooter spectrograph \citep{Vernet2011xshooter} on the Very Large Telescope (VLT), and the Folded-port InfraRed Echellette (FIRE; \citealt{Simcoe2013fire}) spectrograph on the Magellan/Baade telescope,. We summarize the observations in Table~\ref{tab:obs}.

\begin{deluxetable*}{ccccccccccc}
\tablecaption{Summary of Objects Targeted by the Spectroscopic Observations \label{tab:obs}}
\tablewidth{0pt}
\tablehead{
\colhead{ID} & \colhead{RA} & \colhead{Dec} & \colhead{$K$} & \colhead{$z_\mathrm{phot}$} & \colhead{$z_\mathrm{spec}$} & \colhead{Instrument} & \colhead{Obs. Date} & \colhead{Emission Line} & \colhead{Object Name} & \colhead{BL}\\ 
\colhead{} & \colhead{} & \colhead{} & \colhead{[mag]} & \colhead{} & \colhead{} & \colhead{} & \colhead{[UT]} & \colhead{} & \colhead{} & \colhead{}
}
\startdata
37485130234751578 & 02:17:12.223 & $-$04:41:41.336 & $21.98$ & $2.34$ & $2.257$ & NIRES & 2025 Nov 2 & H$\alpha$, \oiii & XMMLSS-z2LRD1 & $\checkmark$ \\
37484979910915093 & 02:17:39.552 & $-$05:08:14.460 & $22.47$ & $2.32$ & $2.101$ & NIRES & 2025 Nov 2/3 & H$\alpha$, \oiii & XMMLSS-z2LRD2 & $\checkmark$ \\
43158859342178868 & 09:59:26.488 & +01:38:46.015 & $22.47$ & $2.38$ & $2.391$ & NIRES & 2025 Nov 2/3 & H$\alpha$, \oiii & ECOSMOS-z2LRD1 & $\checkmark$ \\
41624241757513574 & 23:27:35.881 & $-$00:13:41.654 & $22.09$ & $2.29$ & $2.261$ & FIRE & 2024 Oct 18/19 & 
H$\alpha$, \oiii & DEEP23-z2LRD1 & $\checkmark$ \\
\hline 
37484979910891661 & 02:17:27.071 & $-$05:17:29.330 & $22.04$ & $2.70$ & $2.598$ & X-shooter & 2025 Dec 18 & Ly$\alpha$ & XMMLSS-LAE-1 & $\times$ \\
37484705032989900 & 02:19:11.608 & $-$05:15:09.237 & $21.17$ & $3.41$ & $2.437$ & X-shooter & 2025 Dec 14/15 & H$\alpha$ & XMMLSS-AGN-1 & $\checkmark$ \\
38554250673948704 & 02:20:19.285 & $-$03:49:11.173 & $22.04$ & $2.18$ & $2.350$ & NIRES/X-shooter & 2025 Nov 2, Dec 16 & Ly$\alpha$ & XMMLSS-LAE-2 & $\times$ \\
38553554889236475 & 02:24:24.023 & $-$04:19:29.428 & $22.46$ & $2.51$ & $2.940$ & X-shooter & 2025 Dec 15 & Ly$\alpha$ & XMMLSS-LAE-3 & \nodata \\
37494480378557366 & 02:26:05.816 & $-$04:29:03.077 & $20.02$ & $2.18$ & $2.135$ & NIRES/X-shooter & 2025 Nov 2, Dec 13/14/17& H$\alpha$, Ly$\alpha$ & XMMLSS-AGN2 & $\checkmark$ \\
37494330054695281 & 02:26:44.204 & $-$05:05:33.837 & $22.19$ & $3.16$ & $3.182$ & X-shooter & 2025 Dec 17 & Ly$\alpha$ & XMMLSS-LAE-4 & \nodata \\
\hline
38549161137689328 & 02:18:36.791 & $-$04:09:50.222 & $22.10$ & $2.73$ & \nodata & X-shooter & 2025 Dec 18 & \nodata & XMMLSS-NOLINE-1 & \nodata \\
37484700738020091 & 02:19:11.927 & $-$05:26:24.871 & $22.17$ & $2.92$ & \nodata & X-shooter & 2025 Dec 16 & \nodata & XMMLSS-NOLINE-2 & \nodata \\
37489665720230757 & 02:21:48.858 & $-$04:36:22.026 & $22.37$ & $2.65$ & \nodata & X-shooter & 2025 Dec 16 & \nodata & XMMLSS-NOLINE-3 & \nodata \\
37489528281266493 & 02:22:42.106 & $-$04:40:44.892 & $22.43$ & $3.00$ & \nodata & X-shooter & 2025 Dec 17 & \nodata & XMMLSS-NOLINE-4 & \nodata \\
37489253403359079 & 02:24:13.688 & $-$04:40:10.871 & $21.46$ & $3.02$ & \nodata & X-shooter & 2025 Dec 14 & \nodata & XMMLSS-NOLINE-5 & \nodata \\
37494330054701363 & 02:26:24.854 & $-$05:02:50.966 & $22.20$ & $2.62$ & \nodata & X-shooter & 2025 Dec 15/16 & \nodata & XMMLSS-NOLINE-6 & \nodata \\
37494338644653796 & 02:26:40.103 & $-$04:33:26.584 & $22.30$ & $2.38$ & \nodata & X-shooter & 2025 Dec 16 & \nodata & XMMLSS-NOLINE-7 & \nodata \\
38558498396610385 & 02:27:18.492 & $-$04:24:29.148 & $22.37$ & $2.18$ & \nodata & NIRES/X-shooter & 2025 Nov 3, Dec 15 & \nodata & XMMLSS-NOLINE-8 & \nodata \\
44218380529465803 & 09:55:31.989 & +03:20:11.998 & $22.74$ & $2.83$ & \nodata & X-shooter & 2025 Dec 15 & \nodata & ECOSMOS-NOLINE-1 & \nodata \\
43153636661955138 & 09:57:50.695 & +01:42:45.813 & $22.44$ & $2.11$ & \nodata & X-shooter & 2025 Dec 17 & \nodata & ECOSMOS-NOLINE-2 & \nodata \\
42089738902991461 & 09:59:02.520 & +00:56:54.938 & $22.31$ & $2.53$ & \nodata & X-shooter & 2025 Dec 15 & \nodata & ECOSMOS-NOLINE-3 & \nodata \\
44227726378308581 & 10:04:37.544 & +03:24:02.455 & $21.80$ & $2.58$ & \nodata & X-shooter & 2025 Dec 15 & \nodata & ECOSMOS-NOLINE-4 & \nodata \\
41618877343362187 & 23:26:29.722 & $-$00:24:26.411 & $21.99$ & $2.24$ & \nodata & NIRES & 2025 Nov 3 & \nodata & DEEP23-NOLINE-1 & \nodata \\
41623692001697099 & 23:30:02.175 & $-$00:15:26.987 & $22.67$ & $2.42$ & \nodata & NIRES & 2025 Nov 3 & \nodata & DEEP23-NOLINE-2 & \nodata \\
\enddata
\tablecomments{(1) Catalog ID; (2) Right ascension; (3) Declination; (4) $K$-band magnitude; (5) photometric redshift estimated from the maximum-likelihood solution from \texttt{EAZY}; (6) Spectroscopic redshift measured from our ground-based observation; (7) Spectrograph used; (8) Date of ground-based spectroscopy in UT; (9) Emission line detected; (10) Object name for simplicity; (11) Does the object have a detected broad \ha\ emission line? ``$\cdots$'' means that we do not have spectroscopic coverage of \ha or that we do not detect any emission lines. }
\end{deluxetable*}

\subsection{Targeting Priority}\label{sec:target_selection}

Given the large number of photometrically selected LRD candidates and the limited observing time for spectroscopic follow-up observations, we develop a weighting scheme to prioritize targets with characteristics most similar to LRDs and the feasibility of obtaining spectroscopic observations (i.e., brightness). The formal definition of this weight $w$ is described in detail in Appendix~\ref{app:weight}. Here, we demonstrate in Figure~\ref{fig:weight_visual} how to qualitatively understand the targeting priority in the color-color and color-magnitude spaces. In short, in a first pass, sources with $K<22.5$ are strongly prioritized, since spectroscopically following up fainter objects is challenging due to the bright near-infrared sky. Then, at a given $K$-band magnitude, sources simultaneously exhibiting redder $J-K$ colors, bluer $r-z$ colors, and exhibiting a more pronounced V-shaped SED (quantified by $\Delta_\mathrm{slope}$ as defined in Equation~\ref{eq:dslope}) are prioritized. We assign a weight to each photometric candidate and prioritize spectroscopic targets within each field in descending order of weight.

\subsection{Spectroscopic Observations}\label{sec:observation}

The Keck/NIRES observations (25B-W351, PI: I. Labbe) were carried out on November 2--3 (UT), 2025 under seeing conditions of 0.6--1.3$''$ and with a slit width of $0''.55$ and the correlated double sampling readout mode. For the near-infrared difference imaging, we use an ABC nodding pattern. Each exposure is 300 seconds, and the total exposure time on the Keck targets ranges from 2400--9900 seconds. Telluric standard stars were also observed along the course of the nights, bracketing the science observations. 

The VLT/X-shooter observations (116.294D.001; PI: J. Matthee) were carried out on December 13--18, 2025 (UT) under seeing conditions ranging from 0.5--1.5". We used slit widths of $1''$, $0''.9$, and $0''.9$ in the UVB, VIS, and NIR arms, respectively. We also use a standard ABBA nodding sequence for difference imaging particularly in the NIR. Within a sequence, we took one exposure in the UVB (700 seconds) and VIS (655 seconds) arms and two exposures in the NIR arm ($2\times365$ seconds) at each nodding position. Within a sequence, the total exposure times are 2800, 2620, and 2920 seconds in the UVB, VIS, and NIR arms, respectively. The total exposure times for the VLT targets ranges from 2920 seconds to 8760 seconds. 

The Magellan/FIRE observation was carried out on October 18--19, 2024 (UT) with a slit width of $0''.75$ and the sample-up-the-ramp readout mode. The NIR difference imaging was done in the standard ABBA pattern. The total exposure time for the science frames is 9638 seconds on the single target (DEEP23-z2LRD1; see \citealt{Ma2026z2}). The details of this observation are also described in \cite{Ma2026z2}. 

\subsection{Data Reduction}\label{sec:reduction}

\begin{figure*}[ht!]
    \centering
    \includegraphics[width=\linewidth]{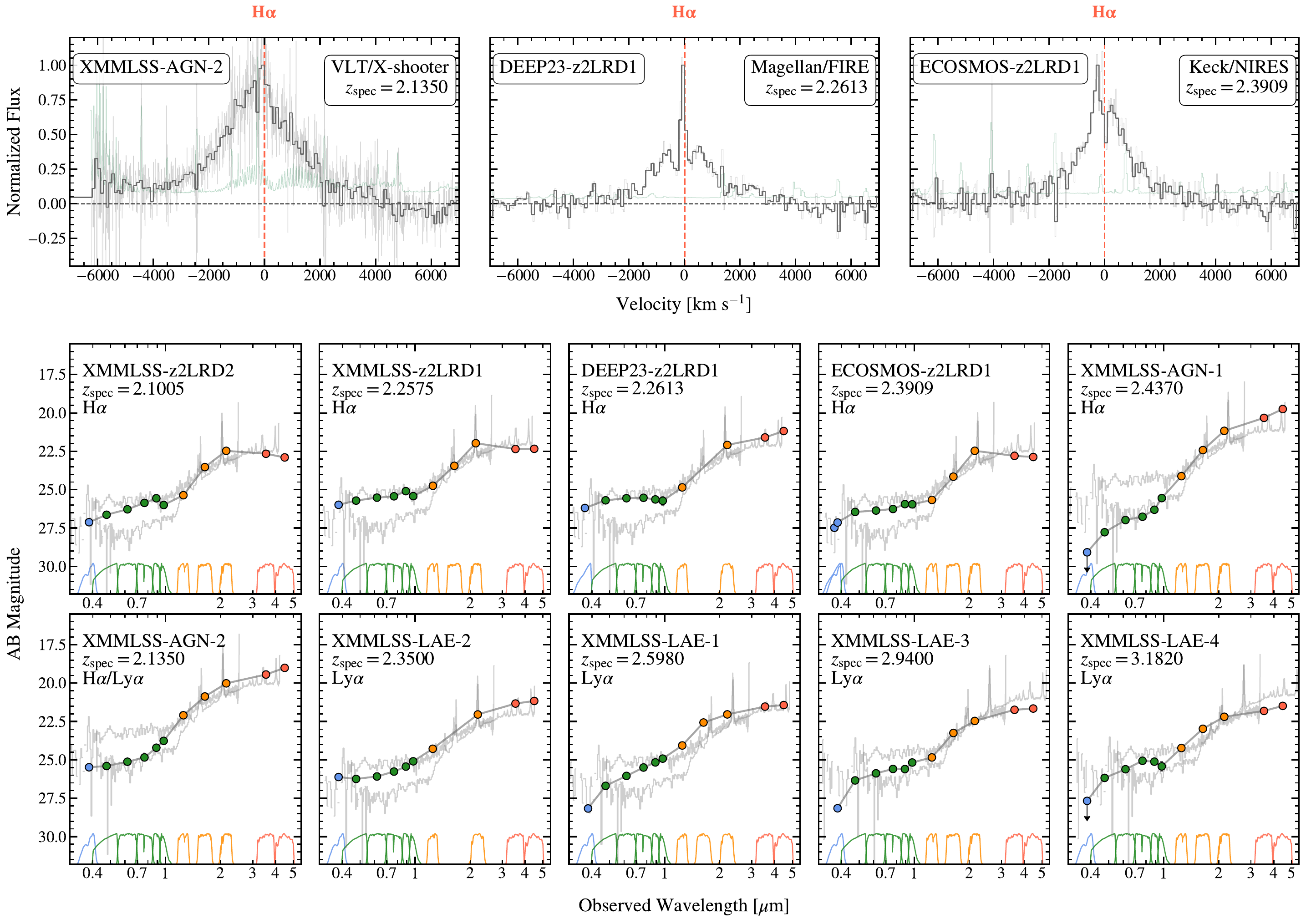}
    \caption{Three examples of the \ha\ emission line among the six spectroscopically confirmed broad-line LRD are shown in top with their photometric SEDs shown on the bottom panels. The light and dark gray colors represent the observed and smoothed (FWHM=7\,\AA) spectra, respectively. The green color represents the uncertainty. Emission lines including \ha\ and the \oiii\ doublet are labeled in objects where they are detected. \textbf{Bottom:} We show photometric SEDs of objects with confirmed spectroscopic redshifts. On the top corner of each panel, we list the hydrogen emission lines detected in these sources. In the background, we overplot the NIRSpec/PRISM spectra of RUBIES-BLAGN-1 \citep{Wang2025brd} and UNCOVER-4286 \citep{Greene2024}, two LRDs identified by JWST. }
    \label{fig:optical_spectra}
\end{figure*}

The Keck/NIRES spectra are reduced with \texttt{PypeIt}, an all-in-one Python-based reduction pipeline for various instruments \citep{Prochaska2020JOSSpypeit, Prochaska2020zenodoPyPeit}. The strong OH emission lines from the sky are used to calibrate the wavelength solution on each individual science frame. The continuum emission is too faint, if present at all, for \texttt{PypeIt} to run the automated trace detection. We therefore identify emission lines (mostly \ha) manually and force the pipeline to extract at the same relative spatial trace in each of the $YJHK$ arms. Manual extraction is performed across all frames using a FWHM that is estimated for each target from a Gaussian fit to the spatial profile of \ha{} in a stack of differenced frame pairs. The individual 2D spectra are coadded into a master 2D spectrum, weighted by the summed inverse variance of pixels within 30\,\AA\ of the \ha\ centroid in the default non-flux-calibrated 1D spectra output by the pipeline. The master 2D spectrum is fed into the pipeline again with a manually placed trace for optimal extraction and the extracted 1D spectrum is flux-calibrated and telluric-corrected with the corresponding telluric standard stars observed on the nights of the observation. 

The VLT/X-shooter spectra are reduced using a combination of the \texttt{EsoRex} libraries\footnote{\href{https://www.eso.org/sci/software/cpl/esorex.html}{https://www.eso.org/sci/software/cpl/esorex.html}} and custom Python routines adapted from the pipeline described by \cite{Matthee2021}. Each observing block (OB) is reduced independently, with an initial flux calibration performed using standard stars observed on the same night. Telluric absorption is corrected with \texttt{molecfit} \citep{Smette2015molecfit}, as incorporated into the \texttt{EsoRex} pipeline, using telluric stars observed on the nights of observing. For each OB, we perform optimal extraction of the one-dimensional spectrum using the spatial profile of \ha\ emission, allowing us to account for changes in seeing and small offsets in slit positioning. Finally, to mitigate differences in slit losses and absolute flux calibration between observations, we rescaled the individual spectra to a common \ha\ flux, defined by the median across all OBs, before combining them through median stacking.

The Magellan/FIRE spectrum was reduced with \texttt{PypeIt} in a similar fashion as the Keck spectra. However, we did not carry out a custom 2D-coadd procedure before extracting a master 1D spectrum as we did for the Keck spectra. Instead, the individual pipeline-produced 1D spectra of each science exposure were flux-calibrated directly and then coadded into a master 1D spectrum. The details of this reduction are described in \cite{Ma2026z2}. 

\subsection{Spectroscopic Redshift and LRD Classification}


\begin{figure}[ht!]
    \centering
    \includegraphics[width=\linewidth]{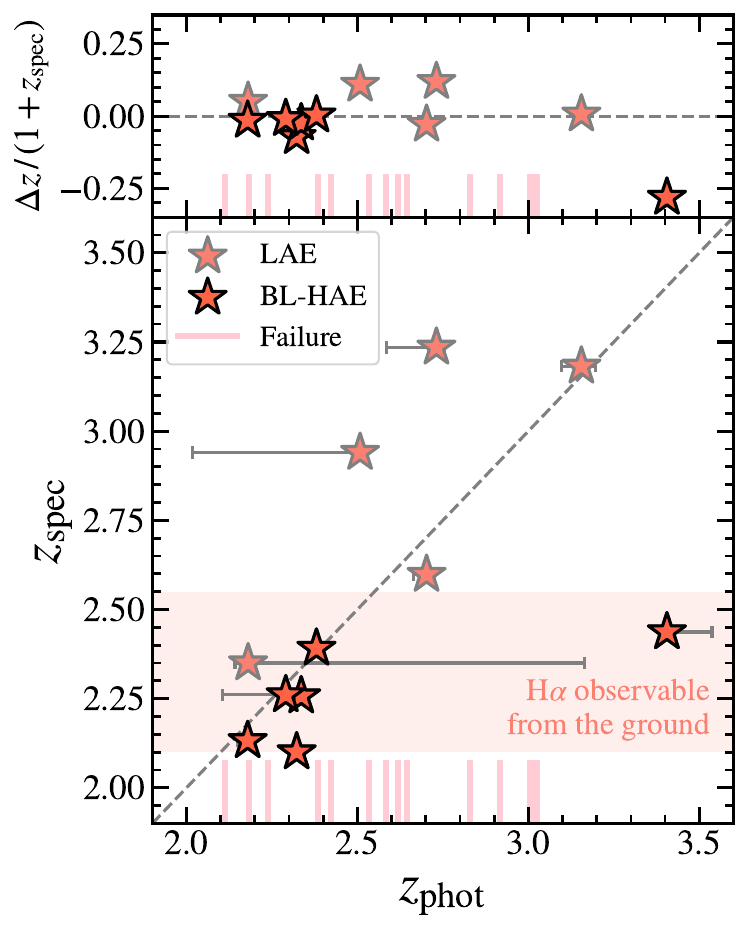}
    \caption{The comparison of photometric redshift and spectroscopic redshifts are shown. Stars with dark edges represent objects with broad \ha\ emission, and those with gray edges represent objects with only \lya\ detected. Objects without any detected emission lines are shown with vertical bars at the median of their photometric redshift estimations. }
    \label{fig:zspec_zphot}
\end{figure}
Among the 24 LRD candidates that we follow up with spectroscopy, we confirm the spectroscopic redshifts of 10 objects from either Ly$\alpha$, \oiii$\lambda\lambda4960,5008$, and/or \ha\ emission lines. Six objects show broad \ha, suggesting AGN/LRD activity in these systems. We show the \ha\ spectra and the photometric SEDs of this spectroscopically confirmed subset in Figure~\ref{fig:optical_spectra}.

The line fitting procedure is described in detail in D. J. Lyon et al. (2026, submitted), so we only outline it briefly here. The Balmer lines are modeled with three components: an exponential profile for the broad emission line (this is commonly seen in many high-redshift LRDs), a narrow Gaussian profile for the narrow emission line, and absorption features modeled as Gaussians (given the limited SNR from these ground-based observation). All components of the Balmer lines are fixed at the same systemic redshift, with the exception of the \ha\ absorption, whose kinematics is untied from the emission features. The \oiii\ doublet is modeled with a single Gaussian component with FWHM fixed to that of the narrow \ha\ emission; the flux ratio of the doublet is assumed to be 2.98. For objects with \lya\ emission, we model the \lya\ with a single Gaussian profile. The continuum is assumed to be constant in the vicinity of each emission line. 

A total of ten objects have a successful spectroscopic redshift measurement, with $2.10\lesssim z_\mathrm{spec}\lesssim3.25$. As shown in Figure~\ref{fig:zspec_zphot}, for sources with $z_\mathrm{spec}\lesssim2.5$ where \ha\ is accessible from the ground, we achieve $|z_\mathrm{phot}-z_\mathrm{spec}|/(1+z_\mathrm{spec})<0.1$. This is because the strong \ha\ emission in LRDs often results in a $K$ band excess in the photometric SED (see Figure~\ref{fig:optical_spectra}). Our modified \texttt{EAZY} code (see Section~\ref{sec:photoz} and Appendix~\ref{app:photoz}), in these cases, would then prefer the LRD template instead of attempting to use extremely dust-reddened templates to account for the excess. Sources above $z_\mathrm{spec}>2.5$ show a larger spread since \ha\ shifts out of the $K$-band, but are largely consistent with $|\Delta z|/(1+z)\lesssim0.1$ as well. The object XMMLSS-AGN-1 is a redshift failure---this is likely due to the fact that \texttt{EAZY} interprets the low-SNR $u^*$ photometry as a drop-out non-detection, which puts the object's photometric redshift at a higher value. 

We detect broad \ha\ emission with FWHM$\sim$1000--2000$\,\mathrm{km\,s^{-1}}$ in six objects, whose \ha\ spectra and photometric SED are shown in Figure~\ref{fig:optical_spectra}. We also detect \ha\ absorption commonly seen in JWST high-redshift LRDs \citep[e.g.,][]{Matthee2026abs} in three out of six broad-line objects, further supporting the potential LRD nature of these broad-line sources. These absorption features exhibit a range of kinematics, from a blueshift of $\sim200\,\mathrm{km\,s^{-1}}$ to a redshift of $\sim40\,\mathrm{km\,s^{-1}}$. 

Five objects also exhibit \lya\ emission. Notably, XMMLSS-AGN-2 shows a broad, double-peaked \lya\ profile alongside broad \ha\ emission. In the remaining sources, two \lya-emitters show no \ha\ detections; for these sources, the lack of continuum constraints prevents us from reliably determining their nature. Nevertheless, our broad \ha\ detection limits of $L_\mathrm{H\alpha}\approx(3$--$5)\times10^{42}\,\mathrm{erg\,s^{-1}}$ at $2.1<z<2.5$ correspond to upper limits on the rest-frame \ha\ equivalent width of $\sim100$--200\,\AA\ at the $K$-band magnitude of these sources ($22.6\lesssim K\lesssim21.9$). These limits are substantially below the typical rest-frame \ha\ equivalent widths of $\sim300$--800\,\AA\ measured for JWST-identified LRDs \citep{Barrufet2026lrdlbd, deGraaff2026sample, Sun2026bhstar}. Thus, if these two sources (with \lya\ but no \ha\ detection) were LRDs with similarly strong \ha\ emission, we would expect to detect their \ha\ lines in our spectroscopic observations given their $K\lesssim22.5$ brightness. Future space-based observations, such as JWST spectroscopy or Roman/HST imaging, will be essential for identifying the contaminating populations in our photometric sample.

Broad \ha\ emission alone is insufficient to securely identify LRDs without full rest-frame UV-to-optical spectroscopy and/or multiwavelength observations, as red quasars and dusty star-forming galaxies can exhibit similar photometric colors \citep[e.g.,][]{Greene2024, Labbe2025, Ma2026z2}. In a companion paper (D. J. Lyon et al. 2026, submitted), we examine archival Chandra data and identify XMMLSS-AGN-1 with $L_\mathrm{2-10\,keV}\approx1.5\times10^{44}\,\mathrm{erg\,s^{-1}}$, whereas most JWST LRDs remain undetected in X-rays \citep[but see][]{Ananna2024, Yue2024, Hviding2026xray}. This source also has a UV slope of $\beta_\mathrm{UV}\approx0.5$, inconsistent with the LRD criterion of \citet{Kocevski2025}, and lacks \ha\ absorption. We therefore classify it as a reddened AGN. Furthermore, D. J. Lyon et al (2026, submitted) find that XMMLSS-AGN-2 is strongly detected ($m_\mathrm{24\,\mu m}=15.73\pm0.03$ AB) in Spitzer/MIPS\,$24\,\mathrm{\mu}$m imaging from the archival Spitzer Enhanced Imaging Products \citep{SEIPdata}. Given the strongly rising mid-infrared emission that does not resemble that of LRDs \citep[e.g.,][]{Setton2025alma} and its UV slope ($\beta_\mathrm{UV}=-0.2$) marginally inconsistent with the \cite{Kocevski2025} criterion, we also classify this object as a reddened AGN. In contrast, the remaining four broad-line sources exhibit UV-optical-near-infrared spectral shapes consistent with JWST LRDs and more than half show \ha\ absorption (D. J. Lyon et al. 2026, submitted); we thus consider these four objects as LRDs and adopt this classification throughout the paper. These four objects are XMMLSS-z2LRD1, XMMLSS-z2LRD2, ECOSMOS-z2LRD1, and DEEP23-z2LRD1. 

\section{Number Density Estimation}\label{sec:analysis}

\subsection{Optical Luminosity Function}\label{sec:olf}

\begin{deluxetable}{ccc}[ht!]
\tablecaption{The binned optical luminosity function of LRDs at $2.10<z<2.55$ \label{tab:lf}}
\tablewidth{0pt}
\tablehead{
\colhead{$M_{5100}$} & \colhead{$\log\Phi$ (Photometry)} & \colhead{$\log\Phi$ (Spectroscopy)} \\ 
\colhead{[mag]} & \colhead{[$\mathrm{cMpc^{-3}mag^{-1}}$]} & \colhead{[$\mathrm{cMpc^{-3}mag^{-1}}$]}
}
\startdata
$-20$ & $-5.65_{-0.21}^{+0.16}$ & $-6.16_{-0.21}^{+0.21}$ \\
$-21$ & $-5.84_{-0.10}^{+0.09}$ & $-6.31_{-0.12}^{+0.11}$ \\
$-22$ & $-6.27_{-0.07}^{+0.07}$ & $-6.75_{-0.11}^{+0.10}$ \\
$-23$ & $-7.55_{-0.30}^{+0.22}$ & $-8.27_{-0.57}^{+0.33}$ \\
$-24$ & $-7.88_{-3.21}^{+0.30}$ & \nodata \\
$-25$ & $<-7.67$ & \nodata
\enddata
\tablecomments{The magnitude bins all have a width of $\pm0.5\,\mathrm{mag}$, and the upper limit is given as the $1\sigma$ value.}
\end{deluxetable}

\begin{figure*}
    \centering
    \includegraphics[width=\linewidth]{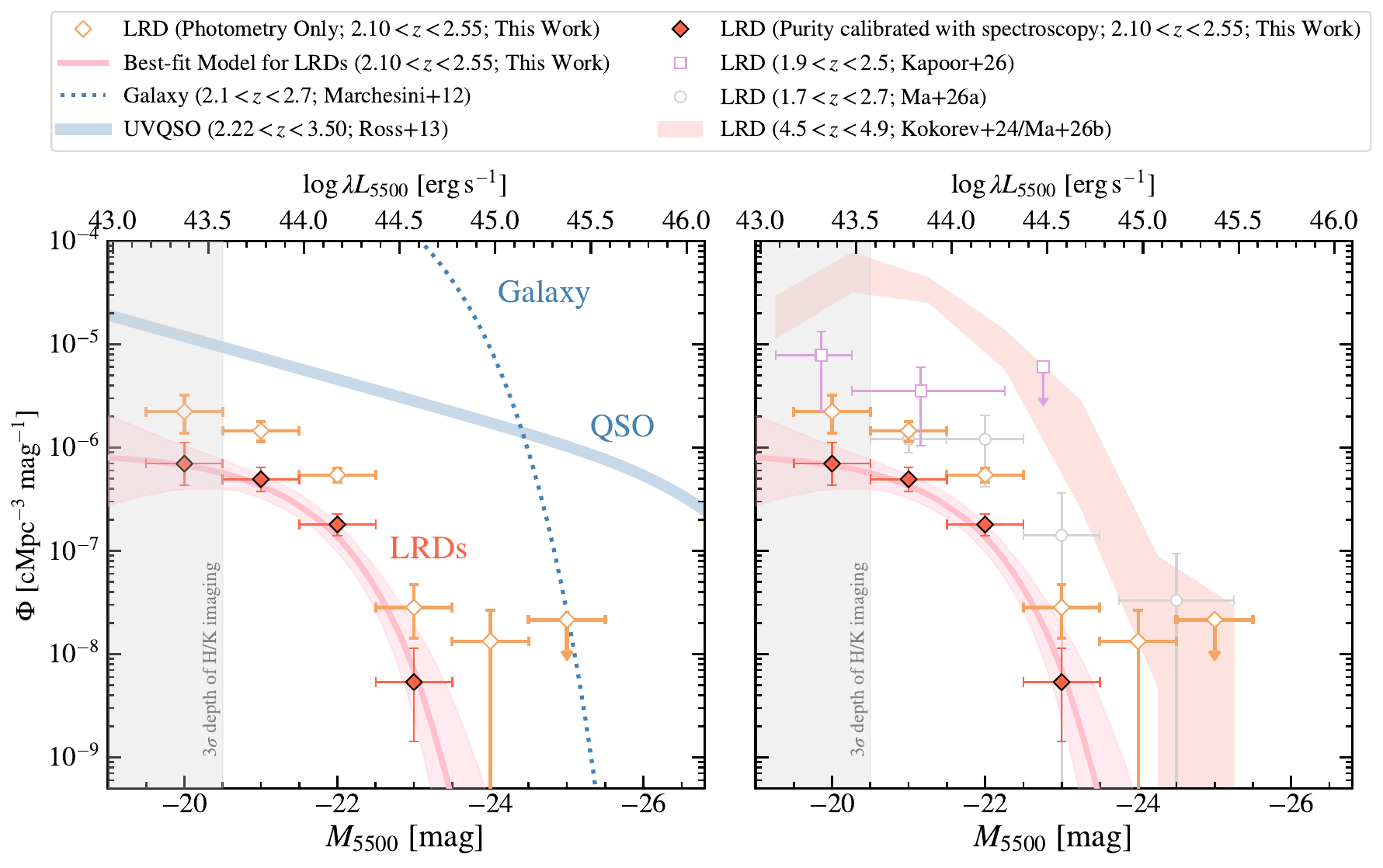}
    \caption{\textbf{Left:} The optical luminosity function of LRDs at $2.10<z<2.55$. The empty and solid diamonds are the luminosity function computed with purely the photometric sample (without spectroscopic calibration; Equation~\ref{eq:phot_lf}) and that computed with the spectroscopic calibration on the purity rate (Equation~\ref{eq:spec_lf}). The gray shaded area roughly corresponds to the $H$ and $K$ $3\sigma$ limit for blind point-source detections at $z\sim2.3$. Although we do detect sources below the limit due to the forced photometry nature of the parent \utoirac\ catalogs (see Section~\ref{sec:data}), the selection is not as robust at those luminosities. The best-fit Schechter function to the spectroscopic luminosity function is also shown along with the $1\sigma$ confidence interval. The quasar luminosity function at $2.22<z<3.5$ \citep{Ross2013QLF_SDSS} is shown as the blue curve. The galaxy luminosity function at $2.1<z<2.7$ \citep{Marchesini2012galaxyLF} is shown as the dotted curve. \textbf{Right:} The optical luminosity functions of LRDs at a range of redshift in the literature \citep{Kokorev2024, Kapoor2026z2, Ma2026cutoff, Ma2026z2} are shown against each other. }
    \label{fig:lf}
\end{figure*}

\begin{deluxetable}{ccc}[ht!]
\tablecaption{Best-Fit Model Parameters for the optical luminosity function of LRDs at $2.10<z<2.55$ \label{tab:bestfit}}
\tablewidth{0pt}
\tablehead{
\colhead{Parameter} & \colhead{Units} & \colhead{Value}}
\startdata
$\Phi^*$ & $\mathrm{{10^{{-7}}\,cMpc^{{-3}}}}$ & {$9.58_{-6.16}^{+5.71}$} \\
$M^*$ & mag & {$-21.3_{-0.8}^{+0.5}$} \\
$\alpha$ & \nodata & {$-0.93_{-0.72}^{+0.74}$} 
\\\hline 
{} & Double Power Law & {} \\\hline 
$\Phi^*$ & $\mathrm{{10^{{-7}}\,cMpc^{{-3}}}}$ & {$2.20_{-1.38}^{+3.33}$} \\
$M^*$ & mag & {$-22.2_{-0.4}^{+0.5}$} \\
$\alpha$ & \nodata & {$-1.56_{-0.49}^{+0.65}$} \\
$\beta$ & \nodata & {$-6.12_{-2.44}^{+2.09}$} \\
\enddata
\tablecomments{\textcolor{black}{$M_{5500}^*=-21.3$}\,mag and \textcolor{black}{$-22.2$}\,mag can be converted into \textcolor{black}{$\lambda L_{5100}=7.8\times10^{43}$\,erg\,s$^{-1}$} and \textcolor{black}{$1.8\times10^{44}$\,erg\,s$^{-1}$}, respectively.}
\end{deluxetable}

Our pilot study \citep{Ma2026z2} uses a subset of this sample in 3-deg$^2$ of the HSC-UltraDeep fields to derive the optical luminosity function of LRDs at cosmic noon. However, that analysis, although demonstrating the significant decline of LRD number density from $z\sim4$ to $z\sim2$, uses a purely photometric sample and therefore suffers from contamination. In this work, we expand our LRD search to nearly 20\,deg$^2$ of the HSC-Deep/UltraDeep fields and use our spectroscopic sample to calibrate the purity of the photometric selection and derive the optical luminosity function of LRDs at $2.10<z<2.55$, where the \ha\ emission lines are accessible from ground-based spectroscopy. Again, the choice of the \textit{optical} luminosity function is motivated by the fact that most of energy output of LRDs is at optical wavelengths \citep{Greene2026lbol, Yanagisawa2026lbollha}. The optical emission is also considered to be directly tracing the power of the central engine of LRDs, whether it is a gas-enshrouded accreting black hole \citep[e.g.,][]{Liu2025break, Sneppen2026} or a supermassive star \citep[e.g.,][]{Chisholm2026gc, Lu2026sms, NandalLoeb2026}. 

We estimate the optical luminosity function using the $1/V_\mathrm{max}$ method \citep{Schmidt1968}. We define the luminosity function in each absolute-magnitude bin as
\begin{equation}\label{eq:spec_lf}
\Phi(M)\Delta M =
\sum_i \left[P_{\mathrm{LRD}}(w_i)\cdot 
\frac{P_{z,i}}{V_{\mathrm{eff},i}}\right]\,\,,
\end{equation}
where the sum runs over all photometrically selected objects in the magnitude bin, regardless of whether they were followed up spectroscopically. Each object is weighted by three factors: the probability that it is an LRD, $P_{\mathrm{LRD}}$ (this is the purity term); the probability that it lies within the redshift interval of interest, $P_z$; and its effective survey volume, $V_{\mathrm{eff}}$. The LRD probability $P_{\mathrm{LRD}}(w_i)$ is defined as the LRD fraction of spectroscopically targeted objects with selection weight $w_i$ (see Section~\ref{sec:target_selection} and Appendix~\ref{app:weight}). For the four objects we consider to be LRDs, $P_\mathrm{LRD}=1$, and the rest of the spectroscopic targets that we classify as non-LRDs, $P_\mathrm{BL}=0$. The redshift probability $P_{z,i}$ gives the probability that the $i$th object lies within $[z_{\min},z_{\max}]=[2.10,2.55]$, and is defined as
\begin{equation}
P_z =
\left\{
\begin{array}{cl}
\displaystyle \int_{z_{\min}}^{z_{\max}}
p(z_{\mathrm{phot}})\,\dd z_{\mathrm{phot}}
& \text{if no } z_{\mathrm{spec}} \text{ is available}, \\[8pt]
1 & \text{if } z_{\min}\leqslant z_{\mathrm{spec}}\leqslant z_{\max}, \\[8pt]
0 & \text{otherwise},
\end{array}
\right.
\end{equation}
where \(p_i(z_{\mathrm{phot}})\) is the photo-$z$ distribution output by \texttt{EAzY}. Finally, the effective volume accessible to the $i$th object is
\begin{equation}
V_{\mathrm{eff},i} = \sum_j \bigg[C_{\mathrm{sel},j}(z_i,M_i)\cdot V_{\mathrm{max},ij}\bigg]\,\,,
\end{equation}
where $j$ indexes the 15 survey subregions over which we estimate the selection completeness $C_{\mathrm{sel},j}$ (see Section~\ref{sec:completeness} and Appendix~\ref{app:completeness}). Here, $V_{\mathrm{max},ij}$ is the maximum comoving volume over which the $i$th object can be detected in subregion $j$. This $V_\mathrm{max}$ is computed using only the detection band (usually HSC/$i$). For nearly all objects in our sample, $V_{\mathrm{max},ij}$ is simply the comoving volume of over $2.10\leqslant z\leqslant2.55$. In the limit where all the objects are photometric candidates (i.e., no purity correction with the spectroscopy), Equation~\ref{eq:spec_lf} is simply reduced to
\begin{equation}\label{eq:phot_lf}
    \Phi(M)\Delta M = \sum_{i}\frac{P_{z,i}}{V_{\mathrm{eff},i}}\,\,.
\end{equation}

The optical luminosity function of the sample is estimated in a Monte-Carlo fashion with 500 realizations. For each realization, we first draw the photometric SED according to the photometric uncertainties. We then draw redshift from the photo-$z$ distribution (constrained within $z_\mathrm{min}\leqslant z_\mathrm{phot}\leqslant z_\mathrm{max}$) for objects not targeted by our spectroscopic campaign and those yielding no $z_\mathrm{spec}$; for those with successful follow-up, $z_\mathrm{spec}$ is used directly. We interpolate the near-infrared photometry to compute $M_{5500}$ (the absolute magnitude at rest-frame 5500\,\AA), which we use along with the redshift to draw the completeness $C_\mathrm{sel}$ in each subregion. The LRD probability is drawn from $P_\mathrm{LRD}\sim B[N_\mathrm{LRD}+1, N_\mathrm{target}-N_\mathrm{BL}+1]$, where $B(x_1, x_2)$ is the Beta function, and $N_\mathrm{LRD}$ and $N_\mathrm{target}$ are the numbers of LRDs and of targeted objects with weight $w_i$, respectively. We also repeat the estimation with Equation~\ref{eq:phot_lf} to obtain the photometric luminosity function without invoking purity calibration from spectroscopy. 

We show the estimated optical luminosity functions in Figure~\ref{fig:lf} and Table~\ref{tab:lf}. We choose to only extend our luminosity function to $M_{5500}\sim-19$ even though the sample contains fainter objects. This is primarily because the $3\sigma$ point source detection depth of the $H$/$K$-imaging corresponds to $M_{5500}\sim -20.5$ at $z=2.3$ \citep{Lawrence2007ukidss, Jarvis2013video}, as we mention in the pilot study \citep{Ma2026z2}. Additionally, at $M_{5500}\lesssim-20.5$, visual inspection also becomes ambiguous, barring us from robustly counting those faint objects. Thus, we consider it sensible to estimate/show the luminosity function only $\sim1\,$mag below the survey depth. 

As shown in Figure~\ref{fig:lf}, without using spectroscopy to calibrate the purity of our sample, the photometric luminosity function we obtain in the full 20-deg$^2$ HSC-D/UD fields is consistent with the one estimated in our pilot study, which only utilizes the 3-deg$^2$ HSC-UD fields \citep{Ma2026z2}.
It also puts the luminosity function of LRDs at $\sim20\times$ lower number density compared to the UV-selected quasars. With our spectroscopic observations calibrating the purity rate of our sample, we obtain another factor of $\sim3$ reduction from the photometric luminosity function. 

Following the treatment and method in \cite{Ma2026cutoff}, we also model the optical luminosity of LRDs as a Schechter function \citep{Schechter1976}, 
\begin{align}
    \Phi(M)\dd M = (0.4\ln10)\times\Phi^*\times10^{0.4(\alpha+1)(M^*-M)} \notag&\\\times\exp\left[-10^{0.4(M^*-M)}\right]\dd M&\,\,,
\end{align}
where $M$ is the absolute magnitude, $M^*$ is the characteristic absolute magnitude determining the position of the bright-end drop-off, $\alpha$ is the faint-end slope, and $\Phi^*$ is the normalization. 
To enable easier comparison with UV-selected quasars, we also model the luminosity function in the form of a double power law \citep{Boyle1988}, 
\begin{equation}
    \Phi(M)\dd M = \frac{\Phi^*\dd M}{10^{0.4(\alpha+1)(M-M^*)}+10^{0.4(\beta+1)(M-M^*)}}\,\,,
\end{equation}
where $\alpha$ and $\beta$ are the faint- and bright-end slopes, respectively---this parameterization is commonly used to describe quasar luminosity functions (QLFs; e.g., \citealt{Kulkarni2019qlf, Ross2013QLF_SDSS}). The best-fit parameters of the Schechter function and the double power law are listed in Table~\ref{tab:bestfit}. It is worth noting that the luminosity function of LRDs drops more steeply towards the bright end than that of the quasars. This may suggest that even at cosmic noon, when quasar activity peaks, LRDs still lack a high-luminosity population, which we discuss further in Section~\ref{sec:cutoff}.

\subsection{Number Density Evolution to Cosmic Noon}\label{sec:number_density}

\begin{figure}[ht]
    \centering
    \includegraphics[width=\linewidth]{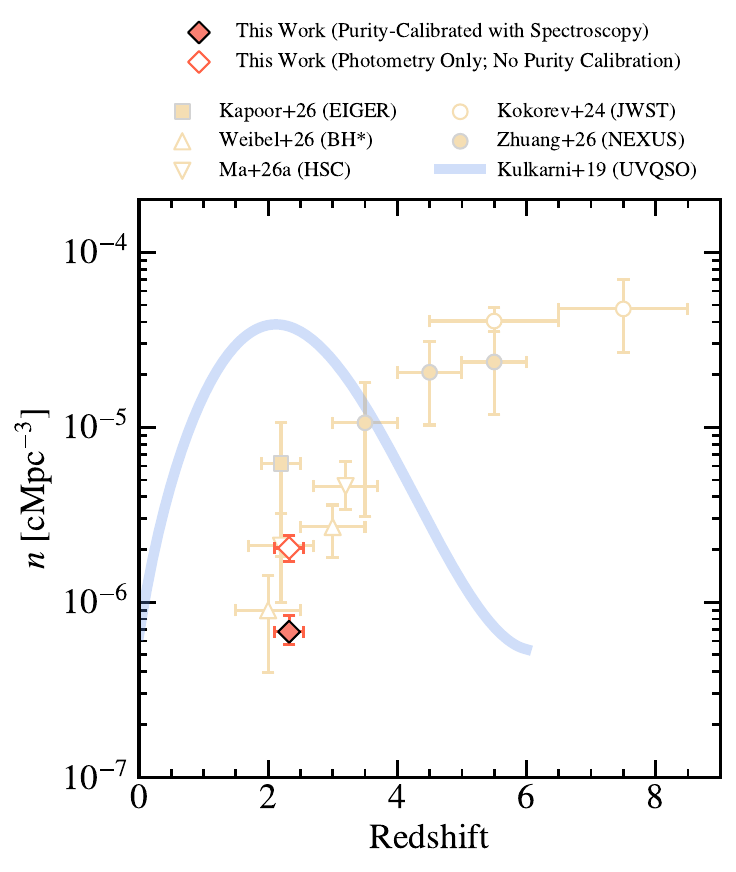}
    \caption{Our photometric (empty red circle) and spectroscopy-calibrated (filled red circle) LRD abundances are shown against LRD abundances at a range of redshifts (yellow points; \citealt{Kokorev2024, Kapoor2026z2, Ma2026z2, Weibel2026bhstar, Zhuang2026nexus}). Estimations with photometric samples are shown as empty markers, and those with spectroscopic samples are shown as filled markers. All LRD number density estimations are integrated down to $M_{5500}=-20.5$ (for estimation USING $M_{5100}$, we apply a $M_{5500}-M_{5100}=-0.26$ correction estimated from our photometric). The quasar abundance evolution \citep{Kulkarni2019qlf} is also shown against that of LRDs. }
    \label{fig:zevolution}
\end{figure}

We integrate the optical luminosity functions at $M_{5500}<-20.5$ and measure the the number density of LRDs at cosmic noon. At $2.10<z<2.55$, our photometric luminosity function returns $2.04_{-0.32}^{+0.35}\times10^{-6}\,\mathrm{cMpc^{-3}}$, consistent with our previous estimation using only photometry within a $\sim6\times$ smaller volume in the HSC/UD fields \citep{Ma2026z2} as shown in Figure~\ref{fig:zevolution}. With the purity calibration from spectroscopy, we yield a factor of $\sim3$ reduction in number density, i.e., $6.8_{-1.08}^{+1.57}\times10^{-7}\,\mathrm{cMpc^{-3}}$. This is nearly a factor of $\sim35$ drop of LRD number density from spectroscopic searches at $z\sim6$ with JWST \citep{Zhuang2026nexus}, when LRDs are at the peak of their abundance. Qualitatively, we (re-)confirm the declining trend of LRD activity from cosmic dawn to cosmic noon that the community has largely agreed upon \citep[e.g.,][]{Inayoshi2025lognormal, Kocevski2025, Kapoor2026z2, Ma2026z2}. Yet, quantitatively, our estimation is lower than most other LRD searches at $z\sim2$. We extensively discuss these differences in Section~\ref{sec:comparison}. 

\section{Discussion}\label{sec:discussion}

\subsection{Comparison with literature}\label{sec:comparison}
There are many works exploring the redshift evolution of LRDs down to cosmic noon \citep{Bisigello2026, Kapoor2026z2, Loiacono2026, Ma2026z2, Weibel2026bhstar}. However, as shown in Figure~\ref{fig:zevolution}, the scatter in the number density estimation is larger than one order of magnitude. Our estimation, with the spectroscopically calibrated purity, is one of the lowest among all. 

First, we examine the caveats in our number density estimation. In particular, it is worth noting that our spectroscopically calibrated number density estimation at $M_{5500}\lesssim-20.5$ is comparable to that of a photometrically selected subset of LRDs with extremely strong breaks \citep{Weibel2026bhstar}. \cite{Weibel2026bhstar} conducted a photometric search of a subset of LRDs at the strong-Balmer--break end using template-fitting methods. At high redshift, they may account for $\sim10\%$ of the LRD population. Thus, at first glance, we would not expect to have comparable number density with their sample. We consider two effects that may be relevant. Firstly, although some sources are supplemented with spectroscopy, \cite{Weibel2026bhstar} select upon photometry, which could introduce interlopers. Secondly, it is very plausible that the completeness is overestimated for our search. The LRD sample we use for completeness estimation inherits the selection/targeting bias of the existing spectroscopic programs; thus, it does not reflect the statistics of the intrinsic distribution of the LRD population. \cite{Hviding2025} have found that color-color selections with JWST filters invoking SNR cuts are only complete at a 50\% level as LRDs with the strongest Balmer breaks are missed due to their faint UV. Similar issues also apply to our selection. Therefore, if the fraction of strong-break LRDs in the spectroscopic sample we use to calibrate completeness is less than the true value, we could still be missing sources even in the completeness-corrected number densities. 

Comparing with other spectroscopically selected LRD samples, our estimation has the largest difference with that of \cite{Kapoor2026z2} (and \citealt{Loiacono2026}, although they only use a subset of the EIGER data), who identify five LRDs over the 140\,arcmin$^2$ of the EIGER footprints using JWST/NIRCam imaging and slitless spectroscopy, although our results agree within $2\sigma$. Yet, as the authors note, they lack full coverage of the rest-frame UV-optical SED and \ha\ emission. The authors infer the LRD nature of their candidates based on their similarity to a combination of LRD features, such as near-infrared sizes, low \hei/Pa\,$\gamma$ ratio, and a lack of X-rays/strong outflows, in addition to the V-shaped SED and compactness criterion. It is possible that contaminating sources still exist in this very small sample that may bring the true number density closer to our estimation. Cosmic variance may also be in play given the small volume searched by these authors. Indeed, the \cite{Kapoor2026z2} search includes all the data analyzed in \cite{Loiacono2026}, who only analyzed the J1030 field (25\,arcmin$^2$) and claim a nearly flat abundance evolution for LRDs, and shows that the J1030 field happens to have largest number of LRDs of all of the EIGER fields. Furthermore, different selection criteria and bolometric corrections in the literature could prevent an easy comparison between our estimations. The choice of bolometric correction depends on the interpretation of the nature of LRDs, which remains open. A consistent source selection, hinging on a common definition of LRDs, is an important step to elucidate the evolution of the number densities and properties of LRDs.  


\subsection{No evolution of the luminosity cutoff of LRDs}\label{sec:cutoff}
\begin{figure}[ht!]
    \centering
    \includegraphics[width=\linewidth]{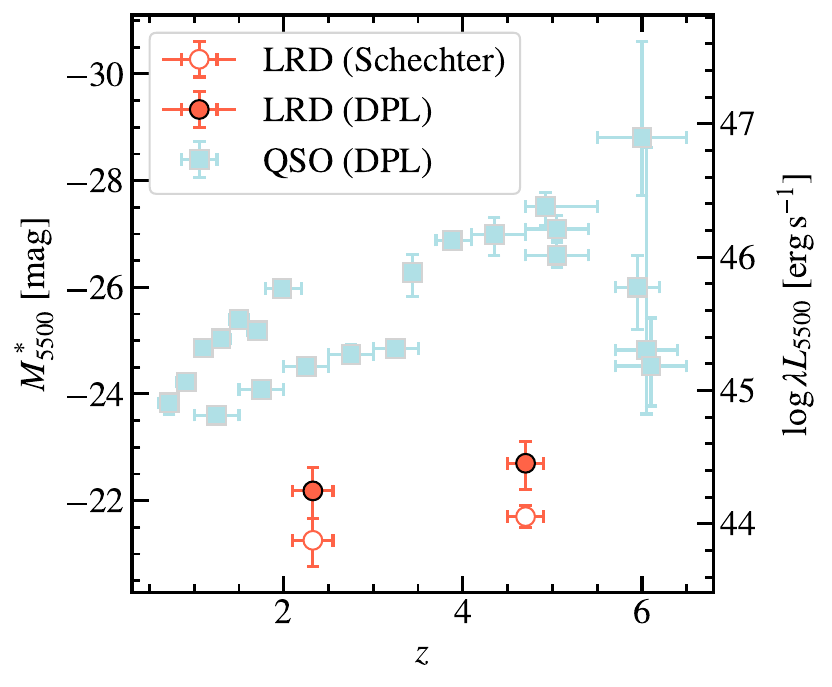}
    \caption{The characteristic absolute magnitude of the best-fit Schechter function and double power law (DPL) to the luminosity function of LRDs is shown as a function of redshift (the $z\sim2$ measurement is from this work, and the $z\sim5$ measurement is from \citealt{Ma2026cutoff}) against that of quasar luminosity functions \citep{Kulkarni2019qlf, Schindler2023panstarrqlf, Yin2026lowzqlf}.}
    \label{fig:cutoff}
\end{figure}

In \cite{Ma2026cutoff}, we identify a sharp cutoff at $M_{5500}\sim-22$ in the optical luminosity function of LRDs at $z\sim5$, with the brightest object we found to be at $M_{5500}\sim-24$. This means that, at $z\sim5$, no LRD is as luminous as the faintest observed quasars. In the left panel of Figure~\ref{fig:lf}, we clearly see that the luminosity function of LRDs at $z\sim2$ falls more steeply towards higher luminosity than the quasar luminosity function does. Furthermore, as shown in the right panel of Figure~\ref{fig:lf}, the shape of the LRD optical luminosity function at $z\sim2$ is very similar to that at $z\sim5$. 

We also show the measured characteristic luminosity, $M^*_{5500}$, of the luminosity functions as a function of redshift in Figure~\ref{fig:cutoff}. Regardless of which functional form we choose to fit the luminosity function, the characteristic optical luminosity of LRDs stay consistently well below those of quasars at both $z\sim5$ and $z\sim2$. We also find that the knee of the optical luminosity function of LRDs does not evolve as a function of redshift at least within our measurement uncertainties, although there is a marginal decreasing trend towards lower redshift. Thus, we may tentatively conclude that only the abundance of LRDs evolves strongly with redshift, but the $M^*$ value of LRD luminosity function does not, in contrast to the quasars. 

If LRDs are indeed powered by accreting black holes, the luminosity cutoff is then mostly determined by a combination of accretion rate and black hole mass. In \citep{Ma2026cutoff}, we interpret the luminosity cutoff as the result of a black hole mass upper bound at $\sim10^7\,M_\odot$ (although degenerate with Eddington ratio and the radiative efficiency) in the underlying black hole mass function of LRDs. Following this argument, our finding in this work suggests that LRDs represent an exclusively low-mass phenomenon across cosmic time. Black hole accretion is scale-free, and therefore a sharp luminosity cutoff is not expected to exist. Yet, this is possible if gas supply cannot keep up with black hole growth, or if the $M_\mathrm{fuel}$--$M_\star$ relation is flatter than the $M_\mathrm{BH}$--$M_\star$ relation. The luminosity cutoff would then naturally lead to a black hole mass upper bound in the black hole mass function of LRDs above which no super-Eddington accretion is sustained since the accretion is fuel-limited. 

Moreover, the persistence of the luminosity cutoff also has implications for models beyond black holes. Theoretical works have used the luminosity cutoff as a strong motivation for the suite of supermassive stars models \citep{Lu2026sms}. In these models, general relativistic instability also naturally determines a maximum mass and luminosity of supermassive stars at $\sim10^{44-45}\,\mathrm{erg\,s^{-1}}$ independent of redshift, from first principles. Therefore, our findings may also partially support these less explored models as well.

\subsection{Why do LRDs disappear towards lower redshift?}

Despite some scatter in measured LRD abundance among different works at cosmic noon (see Section~\ref{sec:comparison}), one thing is certain: LRD abundances does decrease across cosmic time. Searches for LRDs at $z<1$ have reported number densities of $\gtrsim10^{-9}\,\mathrm{cMpc^{-3}}$ \citep{Lin2026desi, Park2026desi}. Although these measurements represent strict lower limits, it is unlikely that these surveys have missed a population larger by four orders of magnitude. Any evolutionary models of LRDs now will have to satisfy four constraints: 
\begin{enumerate}
    \item flat number density evolution at $4<z<8$;
    \item decline of number density from $z\sim4$ to $z\sim2$;
    \item steep cutoff in their luminosity functions;
    \item lack of strong evolution in LRD's cutoff luminosity that we newly identify in this work. 
\end{enumerate}
The last criterion, although subject to large uncertainty, is suggestive that the underlying physical drivers of LRD formation (e.g., black hole mass and Eddington ratio for accretion-powered scenarios, stellar structure for thermonuclear-powered scenarios, or perhaps low metallicity as well) do not evolve significantly over cosmic time. Instead, LRD's decreasing abundance over time reflects that it is the physical conditions required for an LRD to form (or the likelihood of galaxies hosting an LRD event) that becomes rarer.  

\cite{Inayoshi2025lognormal} parameterized the redshift evolution of LRDs with a log-normal function, interpreting the LRD phase as an early episode of gas-enshrouded super-Eddington accretion onto $10^{4-5}\,M_\odot$ black hole seeds. In this case, the stochastic onset of the first accretion episode drives the abundance evolution, while gas depletion eventually terminates the LRD phase. The characteristic black hole mass scale could naturally explain the lack of evolution in the LRD luminosity function shape. Alternatively, \cite{Kapoor2026z2} found that the redshift evolution of LRD abundance resembles the evolution of number of newly formed dark matter halos with masses of $10^{10.5-11.5}\,M_\odot$, suggesting that LRDs may preferentially arise in newly assembled structures that become increasingly rare toward lower redshifts---this a statement independent of the nature of LRD's central engines. The characteristic halo mass scale in this scenario could likewise explain the persistent luminosity function shape. In contrast, \cite{Zhang2026clustering} tentatively suggested that LRDs disappear as their host halos and galaxies grow, diluting their characteristic SED features. However, without a characteristic mass scale, this picture may be difficult to reconcile with the persistent luminosity cutoff identified in this work. Another possibility is that the LRDs favor low-metallicity environments, as suggested by metallicity measurements of some LRDs ($12+\log[\mathrm{O/H}]\lesssim8$ or $\mathrm{[Fe/H]}\sim-2.5$; \citealt{Lin2025sdss, Lin2026desi, Kokorev2026sms}), although the measured value may not trace the gas surrounding the central engine. As galaxies become chemically enriched over time, the fraction satisfying such conditions may decline. Future clustering measurements at $z\sim2$, dedicated metallicity studies, and hydrodynamic simulations will be needed to distinguish these scenarios.

\section{Conclusion}\label{sec:summary}

In summary, we present our wide-field search for LRDs at $1.7<z<3.7$ over 20\,deg$^2$ across the four HSC-D/UD fields. We identify 266 photometrically selected LRD candidates at cosmic noon. Using the Keck, VLT, and Magellan telescopes, we obtained spectroscopic follow-up observations of 24 candidates to confirm their redshifts and search for broad \ha\ emission indicative of their LRD nature. Of these 24 objects, 10 have spectroscopic redshifts confirmed through \lya\ and/or \ha\ emission, and six exhibit broad \ha\ emission, suggesting a tentative purity of $\sim25$--30\% for our photometric sample. A detailed characterization of these broad-line sources is presented in D. J. Lyon et al. (2026, submitted).

Using the spectroscopically calibrated purity, we measure a number density of $6.8\times10^{-7}\,\mathrm{cMpc^{-3}}$ for LRDs with $M_{5500}<-20.5$ at $2.10<z<2.55$, where \ha\ is accessible to ground-based spectroscopy. This confirms the significant decline in LRD abundance from $4<z<8$, where it remains nearly constant, toward later cosmic epochs \citep{Kocevski2025, Kapoor2026z2, Ma2026z2, Zhuang2026nexus}. Moreover, we find that the optical luminosity function of LRDs at cosmic noon declines more steeply toward high luminosities than those of quasars, exhibiting a cutoff at luminosities $\sim50$--100 times fainter than the knee of the quasar luminosity function. A similar result was found at $z\sim5$ \citep{Ma2026cutoff}. Yet, more surprisingly, we find that the knee of the LRD optical luminosity function not only remains at lower luminosities than that of quasars but also shows no significant evolution from $z\sim5$ to $z\sim2$ within our measurement uncertainties. This suggests that an upper mass scale may exist beyond which the central engine, whether a black hole or a supermassive star, can no longer produce the characteristic LRD features. The fact that LRD abundance evolves significantly while the luminosity function knee (and potentially the luminosity function shape) remains unchanged may indicate that the physical conditions required for the LRD phenomenon remain similar across cosmic time but become increasingly rare toward lower redshifts.

We acknowledge that our survey does not provide robust constraints on the evolution of the luminosity function slopes and that our purity estimate is limited by small-number statistics. In the coming years, the Roman Space Telescope and the Rubin Observatory will survey more than 2000\,deg$^2$ of the sky, potentially identifying thousands of LRDs at $1\lesssim z\lesssim2$ and enabling much more robust studies of their luminosity function evolution. These surveys will also provide advantages beyond their sheer large sky coverage. Roman's high angular resolution ($\sim0.1''$; \citealt{Berlfein2025romanpsf}) will enable stringent compactness constraints that are difficult to achieve with seeing-limited ground-based observations, while its slitless spectroscopic capabilities could reduce the need for dedicated spectroscopic follow-up. Together, these upcoming surveys robustly test the evolution of the luminosity function of LRDs across cosmic time. Finally, dedicated JWST spectroscopy of the photometric candidates presented in this work could further constrain the sample purity and enable more robust measurements of LRD abundance at cosmic noon.

\section*{Acknowledgment} 
YM thanks Hanpu Liu, Wenbin Lu, John Chisholm, and Hollis Akins for useful discussion. YM is also grateful for the suggestions from Marta Volonteri that improved the quality of this manuscript. YM, DJL, and IL thank the amazing support from the Keck staff at Waimea, HI during our on-site observing run. YM and AT also thank the extraordinary staff (particularly the chefs!) for their support during our visitor-mode observing run at Paranal Observatory. 

The Hyper Suprime-Cam (HSC) collaboration includes the astronomical communities of Japan and Taiwan, and Princeton University.  The HSC instrumentation and software were developed by the National Astronomical Observatory of Japan (NAOJ), the Kavli Institute for the Physics and Mathematics of the Universe (Kavli IPMU), the University of Tokyo, the High Energy Accelerator Research Organization (KEK), the Academia Sinica Institute for Astronomy and Astrophysics in Taiwan (ASIAA), and Princeton University.  Funding was contributed by the FIRST program from the Japanese Cabinet Office, the Ministry of Education, Culture, Sports, Science and Technology (MEXT), the Japan Society for the Promotion of Science (JSPS), Japan Science and Technology Agency (JST), the Toray Science Foundation, NAOJ, Kavli IPMU, KEK, ASIAA, and Princeton University.

This paper is based in part on data collected at the Subaru Telescope and retrieved from the HSC data archive system, which is operated by Subaru Telescope and Astronomy Data Center (ADC) at NAOJ. Data analysis was in part carried out with the cooperation of Center for Computational Astrophysics (CfCA) at NAOJ.  We are honored and grateful for the opportunity of observing the Universe from Mauna Kea, which has the cultural, historical and natural significance in Hawai`i.

This paper makes use of software developed for Vera C. Rubin Observatory \citep{Juric2017LSSTdatamanagement, Bosch2019lsstpipe, Ivezic2019lsst}. We thank the Rubin Observatory for making their code available as free software at \href{http://pipelines.lsst.io/}{http://pipelines.lsst.io/}. 

This paper also partially makes use of the CLAUDS data \citep{Sawicki2019clauds}, whose data products can be accessed from \href{https://www.clauds.net}{https://www.clauds.net}.

\restartappendixnumbering

\appendix\label{sec:Appendix}

\section{Detailed Breakdown of the Ground-based LRD
Selection}\label{app:sel_breakdown}
In this appendix, we present a detailed breakdown of the number of sources retained at each stage of the selection process described in Section 3.1. Starting from the original \utoirac\ catalog containing 20,485,012 sources, we sequentially apply the quality cuts and LRD selection criteria: 
\begin{enumerate}[start=0]
    \item after applying the quality cuts (e.g., bright star mask), 10,503,279 sources remain;
    \item after applying the signal-to-noise ratio cuts, 4,326,291 sources remain; 
    \item after applying the \texttt{extendedness=0} cut, 275,032 sources remain; 
    \item after applying the $J-K$ cut, 740 sources remain; 
    \item after applying the $r-z$ cut, 568 sources remain; 
    \item after applying the $y-J$ cut, 349 sources remain; 
    \item after applying the $K-[3.6]$ cut, 341 sources remain; 
    \item after visual inspection of the images, 266 sources remain. 
\end{enumerate}
We provide the catalog of the 266 LRD candidates in Table~\ref{tab:all_sample}. 

\begin{deluxetable}{cccccccccccccccc}
\movetabledown=300mm
\tablecaption{The sample of 266 LRD candidates in the HSC-D/UD fields \label{tab:all_sample}}
\tablehead{
    \colhead{ID} & \colhead{RA} & \colhead{Dec} & \colhead{$z_\mathrm{phot}$} & \colhead{$u$} & \colhead{$u^*$} & \colhead{$g$} & \colhead{$r$} & \colhead{$i$} & \colhead{$z$} & \colhead{$y$} & \colhead{$J$} & \colhead{$H$} & \colhead{$K/K_\mathrm{s}$} & \colhead{$[3.6]$} & \colhead{$[4.5]$} \\
    \colhead{} & \colhead{[deg]} & \colhead{[deg]} & \colhead{} & \colhead{[mag]} & \colhead{[mag]} & \colhead{[mag]} & \colhead{[mag]} & \colhead{[mag]} & \colhead{[mag]} & \colhead{[mag]} & \colhead{[mag]} & \colhead{[mag]} & \colhead{[mag]} & \colhead{[mag]} & \colhead{[mag]}
}
\startdata
37494480378557366 & 36.524235 & $-4.484188$ & $2.18$ & \nodata & $25.48$ & $25.40$ & $25.12$ & $24.84$ & $24.21$ & $23.76$ & $22.10$ & $20.87$ & $20.02$ & $19.44$ & $19.01$ \\
37484705032989900 & 34.798367 & $-5.252566$ & $3.41$ & \nodata & $>29.08$ & $27.77$ & $26.98$ & $26.77$ & $26.31$ & $25.55$ & $24.12$ & $22.42$ & $21.17$ & $20.32$ & $19.75$ \\
37489253403359079 & 36.057034 & $-4.669686$ & $3.02$ & \nodata & $>28.92$ & $27.54$ & $26.18$ & $25.99$ & $26.12$ & $25.82$ & $23.97$ & $22.47$ & $21.46$ & $20.55$ & $20.22$ \\
37484700738020091 & 34.799694 & $-5.440242$ & $2.92$ & \nodata & $27.76$ & $26.07$ & $25.48$ & $25.31$ & $25.21$ & $24.95$ & $24.24$ & $22.82$ & $22.17$ & $21.06$ & $20.80$ \\
37489532576226949 & 35.700222 & $-4.529055$ & $3.25$ & \nodata & $>28.75$ & $27.75$ & $25.95$ & $25.76$ & $25.45$ & $25.15$ & $24.75$ & $23.31$ & $22.18$ & \nodata & \nodata \\
37484975615938657 & 34.347679 & $-5.369149$ & $2.39$ & \nodata & $27.32$ & $27.07$ & $26.75$ & $26.32$ & $26.00$ & $25.76$ & $24.65$ & $23.11$ & $22.61$ & $21.62$ & $21.46$ \\
37485125939803712 & 34.134418 & $-4.746638$ & $2.98$ & \nodata & $>29.11$ & $27.57$ & $26.52$ & $25.82$ & $25.63$ & $25.41$ & $24.88$ & $23.18$ & $22.55$ & $22.16$ & $22.08$ \\
37484851061889371 & 34.663750 & $-4.773567$ & $2.37$ & \nodata & $27.93$ & $27.81$ & $27.36$ & $26.73$ & $26.64$ & $26.57$ & $25.08$ & $23.67$ & $22.96$ & $22.69$ & $22.45$ \\
37484842471957967 & 34.642303 & $-5.155046$ & $2.84$ & \nodata & $>29.05$ & $28.58$ & $27.35$ & $27.20$ & $26.85$ & $25.88$ & $25.06$ & $23.70$ & $23.05$ & \nodata & \nodata \\
37485121644812887 & 34.227619 & $-5.099237$ & $3.09$ & \nodata & $>28.99$ & $27.82$ & $27.34$ & $26.64$ & $26.76$ & $26.58$ & $25.55$ & $24.16$ & $23.28$ & $22.44$ & $22.29$ \\
37485121644834606 & 34.300396 & $-4.963159$ & $3.51$ & \nodata & $>29.21$ & $28.09$ & $27.28$ & $27.00$ & $26.48$ & $26.68$ & $25.78$ & $24.84$ & $23.56$ & $23.36$ & $23.03$ \\
38553417450297574 & 36.351879 & $-4.218071$ & $2.76$ & \nodata & $26.68$ & $26.08$ & $24.93$ & $24.28$ & $23.95$ & $23.64$ & $22.80$ & \nodata & $20.76$ & $20.07$ & $19.91$ \\
37484979910891661 & 34.362797 & $-5.291481$ & $2.70$ & \nodata & $28.17$ & $26.69$ & $26.05$ & $25.50$ & $25.17$ & $24.92$ & $24.06$ & $22.57$ & $22.04$ & $21.53$ & $21.43$ \\
38554250673948704 & 35.080353 & $-3.819770$ & $2.18$ & \nodata & $26.12$ & $26.24$ & $26.08$ & $25.76$ & $25.44$ & $25.10$ & $24.28$ & \nodata & $22.04$ & $21.33$ & $21.16$ \\
38549161137689328 & 34.653297 & $-4.163950$ & $2.73$ & \nodata & $>27.80$ & $26.59$ & $25.63$ & $25.16$ & $24.89$ & $24.59$ & $24.14$ & \nodata & $22.10$ & \nodata & \nodata \\
37494330054701363 & 36.603558 & $-5.047491$ & $2.62$ & \nodata & $>27.67$ & $26.94$ & $26.66$ & $26.51$ & $25.89$ & $25.34$ & $24.78$ & \nodata & $22.20$ & $21.13$ & $20.93$ \\
38558365252620620 & 37.079536 & $-4.249302$ & $2.57$ & \nodata & $>27.60$ & $27.64$ & $26.95$ & $26.21$ & $25.99$ & $25.61$ & $24.27$ & \nodata & $22.23$ & $21.85$ & $21.74$ \\
37494338644653796 & 36.667095 & $-4.557384$ & $2.38$ & \nodata & $>28.76$ & $>28.34$ & $27.14$ & $26.85$ & $26.37$ & $25.59$ & $24.41$ & \nodata & $22.30$ & $22.19$ & $22.12$ \\
37489665720230757 & 35.453575 & $-4.606118$ & $2.65$ & \nodata & $>28.67$ & $27.69$ & $27.35$ & $26.78$ & $26.66$ & $25.77$ & $24.78$ & \nodata & $22.37$ & \nodata & \nodata \\
38558498396610385 & 36.827049 & $-4.408097$ & $2.18$ & \nodata & $26.78$ & $26.54$ & $25.98$ & $25.54$ & $25.51$ & $25.00$ & $24.77$ & \nodata & $22.37$ & $21.88$ & $21.83$ \\
38553834062108772 & 35.711540 & $-4.177325$ & $2.34$ & \nodata & $>28.37$ & $28.12$ & $26.83$ & $27.14$ & $25.86$ & $25.18$ & $24.43$ & \nodata & $22.38$ & $21.87$ & $21.84$ \\
38554121824917521 & 35.320137 & $-3.600252$ & $3.44$ & \nodata & $>28.54$ & $27.10$ & $26.30$ & $25.54$ & $25.55$ & $24.99$ & $24.73$ & \nodata & $22.51$ & $21.87$ & $21.77$ \\
37485121644833863 & 34.153954 & $-4.965586$ & $2.46$ & \nodata & $>29.09$ & $27.84$ & $26.97$ & $26.67$ & $26.44$ & $26.06$ & $24.53$ & $23.06$ & $22.51$ & $21.83$ & $21.70$ \\
37494196910714418 & 36.862579 & $-4.888385$ & $2.62$ & \nodata & $28.62$ & $27.06$ & $26.40$ & $26.43$ & $25.52$ & $24.94$ & $24.77$ & \nodata & $22.55$ & $21.58$ & $21.39$ \\
38553984385973914 & 35.443233 & $-3.471154$ & $2.75$ & \nodata & $>27.59$ & $27.12$ & $26.49$ & $26.02$ & $25.67$ & $25.06$ & $24.70$ & \nodata & $22.56$ & \nodata & \nodata \\
\vdots & \vdots & \vdots & \vdots & \vdots & \vdots & \vdots & \vdots & \vdots & \vdots & \vdots & \vdots & \vdots & \vdots & \vdots & \vdots 
\enddata
\tablecomments{Limits are given as $3\sigma$, and ``$\cdots$'' denotes unavailable photometry. The full table will be available online. }
\end{deluxetable}

\section{Photometric Redshift Estimation}\label{app:photoz}

\begin{figure}[h]
    \centering
    \includegraphics[width=\linewidth]{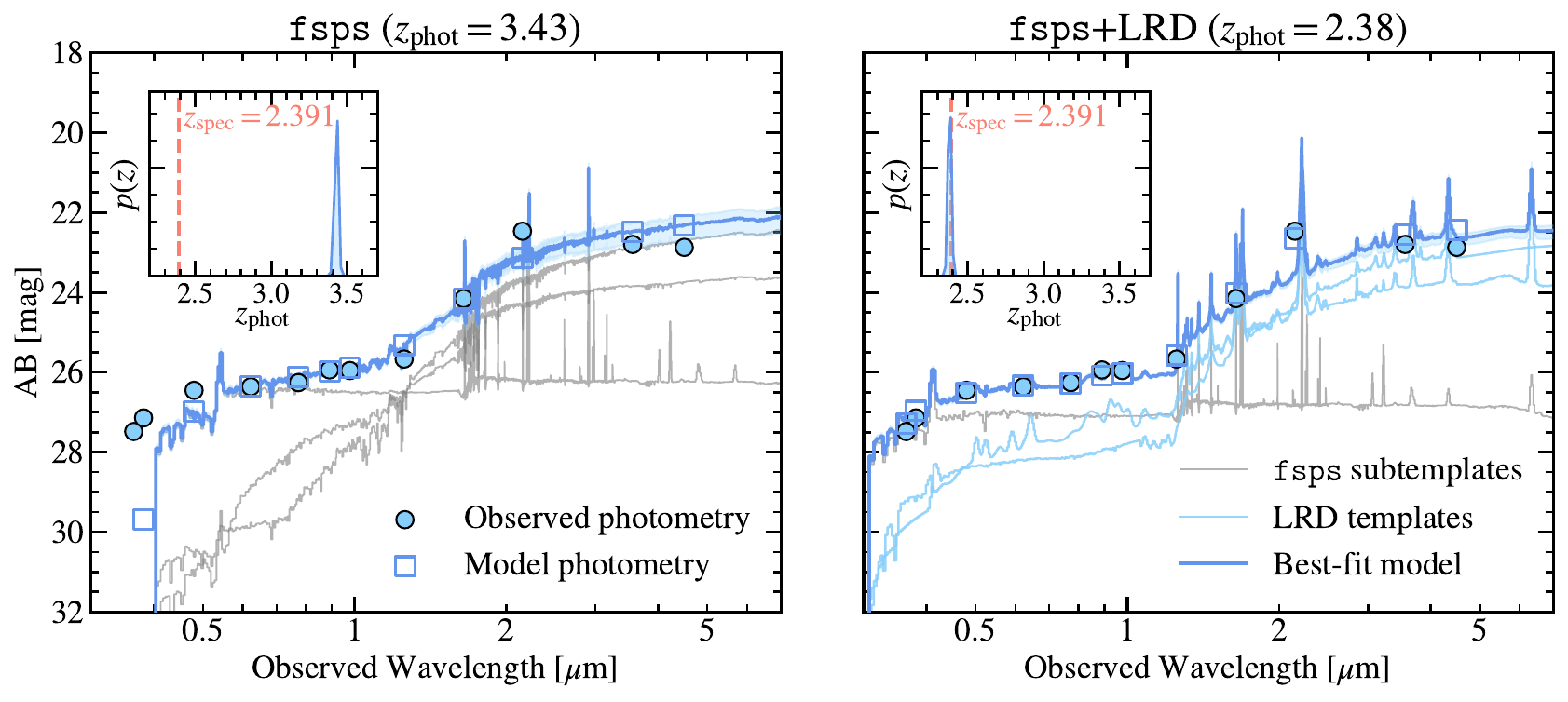}
    \caption{\textbf{Left:} The best-fit \texttt{EAZY} model using the default template suite is shown against the photometry of ECOSMOS-z2LRD1. The output photometric redshift is significantly higher than the spectroscopic redshift. \textbf{Right:} The best best-fit \texttt{EAZY} model using the LRD-supplemented template suite is shown against the photometry of ECOSMOS-z2LRD1. The photometric redshift is now consistent with the spectroscopic redshift. }
    \label{fig:photo_z_comparison}
\end{figure}

Photometric redshifts are estimated using \texttt{EAZY} \citep{Brammer2008eazy}. As described in Section~\ref{sec:photoz}, we supplement the default \texttt{tweak\_fsps\_QSF\_12\_v3} template suite with two LRD templates. We illustrate the effect of these additional templates in Figure~\ref{fig:photo_z_comparison} using ECOSMOS-z2LRD1 as an example. The strong \ha\ emission in LRDs can produce a $K$-band excess, which, in the absence of LRD templates, \texttt{EAZY} attempts to reproduce using the continuum emission of extremely red templates, biasing the photometric redshifts toward higher values. With the inclusion of LRD templates, the $K$-band excess can instead be correctly attributed to strong \ha\ emission, resulting in more accurate photometric redshift estimates.

\section{Completeness in Individual Fields}\label{app:completeness}
\begin{figure}[h]
    \centering
    \includegraphics[width=\linewidth]{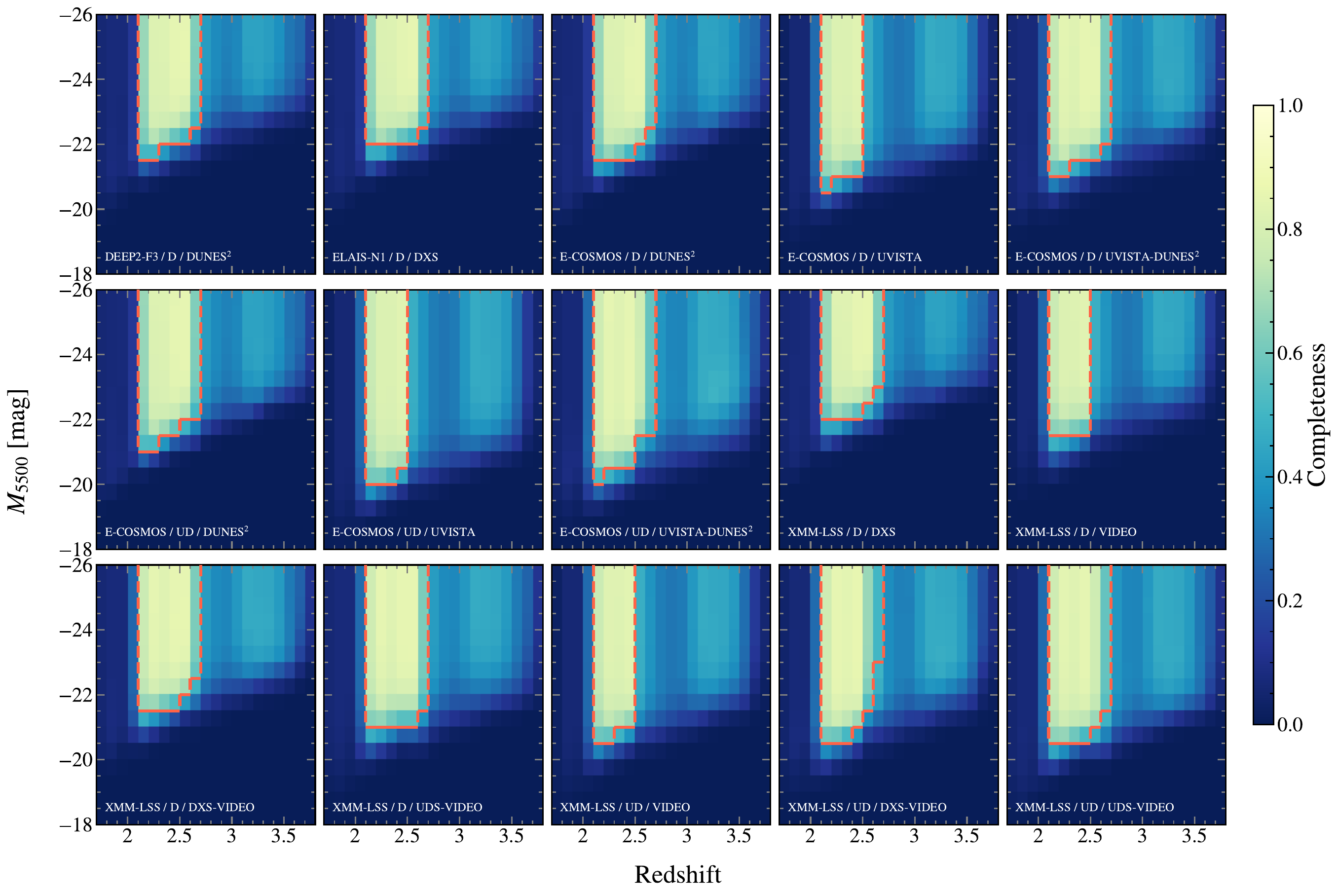}
    \caption{The completeness of each subregions of the HSC D/UD fields (depending on the different combination of imaging depth i the optical and near-infrared) is shown as a function of redshift and optical luminosity. The red dashed lines show the parameter spaces where our selections is 50\% complete. }
    \label{fig:completeness_multi}
\end{figure}

We show the estimated selection completeness in each of the four HSC-D/UD in Figure~\ref{fig:completeness_multi}. Since the NIR imaging in the E-COSMOS and XMM-LSS fields is from multiple surveys with different footprints and depths, the estimated completeness vary slightly across them. 

\section{Targeting Priority for Spectroscopy}\label{app:weight}
In Section~\ref{sec:target_selection}, we qualitatively describe the weighting schemed that we adopt to assign spectroscopic targeting priority to the photometric candidates. In this appendix, we define this weight $w$ in detail as 
\begin{equation}\label{eq:weight}
w =
\frac{
\mathcal{N}(J-K;2,3)
+
\mathcal{N}\!\left[z-r;-1,1\right]
+
\mathcal{N}(\Delta_{\mathrm{slope}};1,4)
}{
1+\exp\!\left[2(K-22.5)\right]
}\,\,
\end{equation}
where
\begin{equation}
\mathcal{N}(x;a,b)
=
\frac{\min\!\left[\max(x,a),b\right]-a}{b-a}.
\end{equation}
and $\Delta_\mathrm{slope}$ is the difference between the optical and UV slopes defined as 
\begin{equation}\label{eq:dslope}
\Delta_{\mathrm{slope}}
=
\frac{\log\!\left(f_{K}/f_{J}\right)}
     {\log\!\left(\lambda_{K}/\lambda_{J}\right)}
-
\frac{\log\!\left(f_{z}/f_{r}\right)}
     {\log\!\left(\lambda_{z}/\lambda_{r}\right)},
\end{equation}
where $f_x$ and $\lambda_x$ are the flux and effective wavelength of the $x$-band. Equation~\ref{eq:weight} can be understood intuitively. The sigmoid function on the $K$-band magnitude sets a strong priority on the K-band brightness threshold below which efficient spectroscopy cannot be carried out due to the bright 2\,$\mathrm{\mu m}$ sky background---brighter objects are prioritized. Then, sources simultaneously showing redder optical ($J-K$) colors, bluer UV colors ($z-r$), and whose SED is more V-shaped ($\Delta_\mathrm{slope}$) are prioritized at fixed $K$-band brightness.



\bibliography{ref}{}
\bibliographystyle{aasjournal}

\end{CJK*}
\end{document}